\documentclass[ascmo, manuscript]{copernicus}
\begin{document}
\nolinenumbers
\title{Non-stationary time series attribution for heatwaves over Europe}
 
 
\Author[1][pmeurer@uni-bonn.de]{Pascal}{Meurer} 
\Author[1]{Sebastian}{Buschow}
\Author[1]{Svenja}{Szemkus}
\Author[1][pfried@uni-bonn.de]{Petra}{Friederichs}
 
\affil[1]{Institute of Geosciences, University of Bonn, Auf dem H\"ugel 20, 53121 Bonn, Germany}
 
\runningtitle{Non-stationary time series attribution for heatwaves over Europe}
 
\runningauthor{Meurer et al.}

\received{}
\pubdiscuss{} 
\revised{}
\accepted{}
\published{}
 

\firstpage{1}
\maketitle
\begin{abstract}
The increasing occurrence of extreme weather events since the beginning of the 21st century has led to the development of new methods to attribute extreme events to anthropogenic climate change. The way in which the extreme event is defined has a major influence on the attribution result. A frequently overlooked aspect concerns the temporal dependence of extremes.
This study presents an approach for attributing complete time series during extreme events to anthropogenic forcing. The approach is based on a non-stationary Markov process using bivariate extreme value theory to model the temporal dependence of the time series. We calculate the likelihood ratio of an observational time series from ERA5 given the distributions as estimated from CMIP6 simulations with historical natural-only and natural and anthropogenic forcing scenarios. The spatial fields are condensed by the extremal pattern index (EPI) as a compact description of spatial extremes. In addition, the study examines the extent to which attribution statements about the occurrence of extreme heat events change when the effect of the mean warming is eliminated. The resulting attribution statement provides very strong evidence for the scenario with anthropogenic drivers over Europe, especially since the beginning of the 21st century. For central and southern Europe, the influence of anthropogenic greenhouse gas emissions on heatwaves could already have been proven in the 1960s using today's knowledge. There is no reliable signal apart from a general shift in the temperature distribution, neither in terms of the temporal dependence of extreme heat days nor in terms of the shape of the extreme value distribution.
\end{abstract}

\copyrightstatement{TEXT}
 
\introduction  
 
The considerable increase in greenhouse gases resulting from the extensive use of fossil energy sources by humans, coupled with the substantial alterations in land use, has led to the emergence of an ongoing global climate change phenomenon, which began approximately a century ago \citep{IPCC_2021_WGI_Ch_2}.
In recent years, annual greenhouse gas emissions have reached unprecedented levels \citep{IEA2021}. This trend is mirrored by the concentrations of $\text{CO}_2$, $\text{CH}_4$ and $\text{N}_2\text{O}$, which have reached levels that have not been seen for at least 800,000 years \citep{IPCC_2021_WGI_Ch_2}. The Earth's climate system is undergoing a transformation as a consequence of alterations in atmospheric composition. In the 1980s and 1990s, scientific interest therefore focused on detecting observed changes in the climate system (e.g.~the increase in global mean temperature). These detection studies have already demonstrated the strong signal of anthropogenic climate change in observed changes in global mean surface temperature \citep{Hegerl2006}.
 
In addition to the increase in global mean temperature, the occurrence of extreme events such as heatwaves has increased since the beginning of the 21st century. This has led to a growing number of studies on the attribution of extreme weather events \citep{hulme_attributing_2014,IPCC_2021_WGI_Ch_11}. One of the first studies of this kind was conducted by \citet{stott_2003}.
Such studies aim to investigate whether changes in extreme weather can be attributed to anthropogenic climate change. Several methods for the attribution of extreme events have been established that focus the attribution statement on a single extreme event, such as a heatwave or flood event \citep{45935409b24941868afbc5d93c60d166}. Typically, a data compression method (e.g., average over a region) is used to attribute the event based on univariate extreme value statistics \citep{Philip2020}.
The attribution is then performed by calculating the probability of the event given factual and counterfactual conditions, based on the so-called causal counterfactual theory \citep{Hannart2016}. These probabilities are then compared using a probability ratio, or in the case of a time series the likelihood ratio \citep{Seong2022}.
 
The temporal evolution and dependence are often ignored, or techniques are used to break down the time series into individual clusters of extreme values \citep{Philip2020, SIPPEL201525, TheDeadlyCombinationofHeatandHumidityinIndiaandPakistaninSummer2015}.
A useful extension in extreme value statistics is the modelling of temporal dependence for extreme events \citep{Markov_model_wind}. We use this approach to include temporal dependence in the attribution approach.
The question of attribution for periods without extreme events is also relevant. For example, one may consider the attribution of a summer with cold spells and heat extremes. The European summer of 2025, for example, included cold spells as well as heat extremes (see Sect.~\ref{sec:example2025}). In this study, we present an approach for attributing complete time series that consist of extreme and non-extreme events. This allows us to extend the attribution not only to entire summer periods, but also to a series of summers. One can then assess whether non-extreme parts of the time series, in which a high threshold is not exceeded, contribute to evidence for the greenhouse gas scenario, and if not, how strong the opposing evidence is. In this sense, the likelihood ratio can be updated by adding additional observations to the attribution.
 
Since we want to process data with spatial and temporal dimensions and obtain reliable probabilities for extreme events, the first step is to reduce the dimensions in space using a suitable data compression method. We use a method presented by \citet{Szemkus2024} based on a decomposition of extremal dependence. Their approach is closely related to principal component analysis, as used, for example, to identify teleconnections in the atmosphere, and follows \citet{Cooley2019} and \citet{jiang_principal_2020}. Instead of correlation or covariance, a measure specifically designed for the dependence of extremes is used to identify spatially coherent patterns. The extremal pattern index (EPI) as introduced in \citet{Szemkus2024} then provides a spatially aggregated measure of the strength of an extremal pattern of a meteorological variable.

We can use the standard peaks-over-threshold (POT) approach to model extreme values of the EPI above a high threshold. In this approach, threshold exceedances follow a generalized Pareto distribution (GPD). As multiple threshold exceedances can occur in succession, the times of their occurrence cannot be treated as independent events. Therefore, a common strategy is to decluster the time series and retain only the cluster maximum. Although this reduces serial dependence among exceedances, it introduces additional choices; for example, the declustering rule \citep[Sect.~10.4]{Beirlant}. 
If a general warming trend is present, the number of threshold exceedances and, therefore, the clustering, will change. In this case, the clustering must also be made time-dependent; that is to say, different cluster lengths in different years must be considered. Alternatively, we can model the time-dependence structure directly. This enables us to model the temporal dependence and consider a changing heatwave definition by increasing the threshold.

The temporal dependence of the daily EPI time series is considered using a Markov process. The Markov process can be described by an approximate likelihood based on bivariate extreme value theory \citep{Smith_markov, Beirlant}. This likelihood is based on the censored likelihood model, which divides the two-dimensional plane into four regions according to whether or not the variables exceed the threshold value. The corresponding two-dimensional variable represents two consecutive observations in the Markov process. The resulting dependence is thus analogous to the autocorrelation in a time series. Declustering the extremes in a time series is therefore no longer necessary.
 
Here, we focus on a parametric model of temporal extremal dependence formulated within the class of asymptotically dependent bivariate extreme-value models. This assumption is consistent with the pre-asymptotic nature of the considered data, in which exceedances occur at high but finite thresholds. However, recent studies suggested models that can also represent asymptotic independence, in which extremal dependence vanishes in the limit but may still be present at finite levels \citep[e.g.,][]{ledford_tawn2,ramos_ledford,Wadsworth}. We do not pursue this extension here, but we regard it as an important area for future research.

The approach is applied to different scenarios, represented by a multi-model ensemble of CMIP6 simulations, so that we can compare the likelihood of an observational time series given different scenarios.
For each scenario and climate model, a set of model parameters can be estimated, and an attribution statement can be made using the respective likelihood ratio.

This study examines heatwaves, which are described by the maximum daily temperature near the surface. We would like to answer the following questions in particular.
\begin{enumerate}
    \item Can [changes in] the temporal evolution of a heatwave be attributed to anthropogenic emissions? 
    \item Is there a climate change signal in the tail behaviour of heatwave extremes beyond a general shift in the temperature distribution? 
\end{enumerate}
For both questions, we adapt the Markov process model further to account for non-stationary conditions and ongoing changes in the climate system, following \citet{Hannart2016}. They pointed out that the stationarity assumption is unrealistic because the mean temperature and thus the extremes have changed over time. Based on slowly varying covariates, possible changes in the parameters can be modelled.
 
The article is structured as follows. Section \ref{sec:data} introduces the data used in this study, as well as the compact description of multivariate weather events in terms of EPI. The daily maximum surface temperatures in Europe in the summer of 2025 are used to illustrate the methods. In Sect.~\ref{sec:theory} we briefly introduce the attribution approach for extreme events, present the derivation of the approximate likelihood for the Markov process, and introduce non-stationarity. In Sect.~\ref{sec:attribution_workflow}, the workflow for the attribution is presented using the Markov process model. In Sect.~\ref{sec:results}, the results for heatwaves are presented. Lastly, Sect.~\ref{sec:conclusion} discusses these results and gives our concluding remarks.
 
\section{Data and Indices}\label{sec:data}
\subsection{Data}
\label{sec:data_part}
We used the daily maximum 2-meter temperature (T2max) of the ERA5 reanalysis \citep{Hersbach2020} in Europe. ERA5 provides a spatially and temporally consistent description of the atmospheric state since 1940 based on data assimilation. We focus our study on daily time series during the northern hemisphere summer seasons June to August (JJA).
 
\begin{figure*}[t]
    \includegraphics[width=1\textwidth]{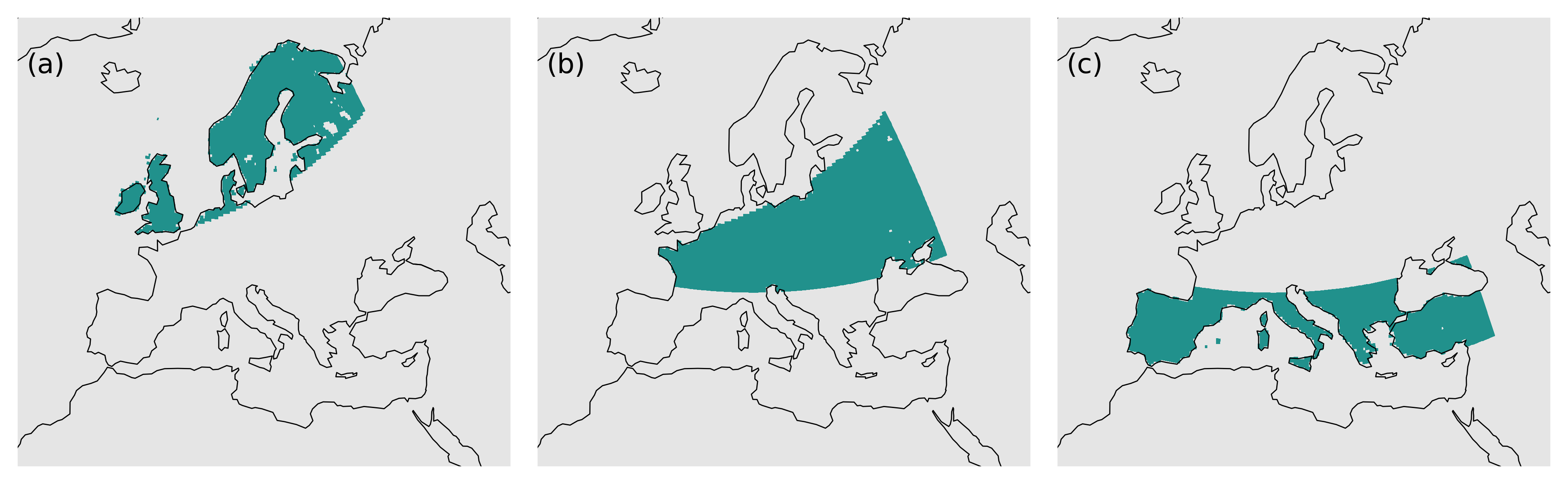}
    \caption{AR6 regions \citep{regions} of (a) northern, (b) central, and (c) southern Europe as used in this study with North Africa excluded and restricted to land points only. A grid cell is considered a land point when at least $50~\%$ of the cell is occupied by land.}\label{fig:landmask}
\end{figure*}

The climate simulations are taken from the CMIP6 and DAMIP v1.0 projects \citep{Gillett2016}. The scenarios we use are historical-natural (HIST-NAT) without anthropogenic forcing, the historical scenario (HIST) and SSP2-4.5 with anthropogenic forcing.
The respective circulation model, the variable, and the number of members available in each ensemble are given in Table~\ref{tab:cmip6}. As for the ERA5 dataset, we focus on JJA months. The study area consists of only land points. A grid cell is considered a land point when at least $50~\%$ of the cell is occupied by land. We used the IPCC AR6 regions shown in Fig.~\ref{fig:landmask} of northern, central, and southern Europe \citep{regions}. The regions in North Africa are excluded from the analysis due to their different climates with extremely dry desert areas.
 
\begin{table}[!ht]
    \caption{Number and type of CMIP6 simulations with HIST (1940--2014), SSP2-4.5 (2015--2025), and HIST-NAT (1940--2020) for T2max used in this study.}\label{tab:cmip6}
    \small
    \begin{tabular}{ |p{2.9cm}||p{0.85cm}||p{0.85cm}||p{1cm}|}
        \hline
        \multicolumn{4}{|c|}{Number and type of CMIP6 simulations} \\
         \hline
        Variable& T2max & T2max  & T2max  \\
        Scenario & HIST  & SSP2-4.5 & HIST-NAT  \\
        \hline
        ACCESS-ESM1-5 & 19 & 18 &  3 \\
        BCC-CSM2-MR & 3 & 1 & 3 \\
        CanESM5 & 50 & 50 & 10   \\
        CNRM-CM6-1 & 30 & 1 & 3  \\
        HadGEM3-GC31-LL & 4 & 4 & 4   \\
        IPSL-CM6A-LR & 32 & 11 &  6   \\
        MIROC6 & 50 & 49 &  50   \\
        MPI-ESM1-2-LR & 50 & 50  & 30   \\
        MRI-ESM2-0 & 7 & 4 &  5  \\
    \hline
    \end{tabular}
\end{table}
 
\subsection{Extremal pattern index (EPI)}\label{sec:EPI}
 
Our compact representation of a spatially extended multivariate weather event uses the EPI as suggested in \citet{Szemkus2024}. The approach is based on the decomposition of high-dimensional data based on a description of the tail dependence within the framework of regular variation \citep{Cooley2019}. The reader is referred to \citet{Szemkus2024} for the details of the method.
 
\begin{figure}[h]
    \centering
    \includegraphics[width=0.55\linewidth]{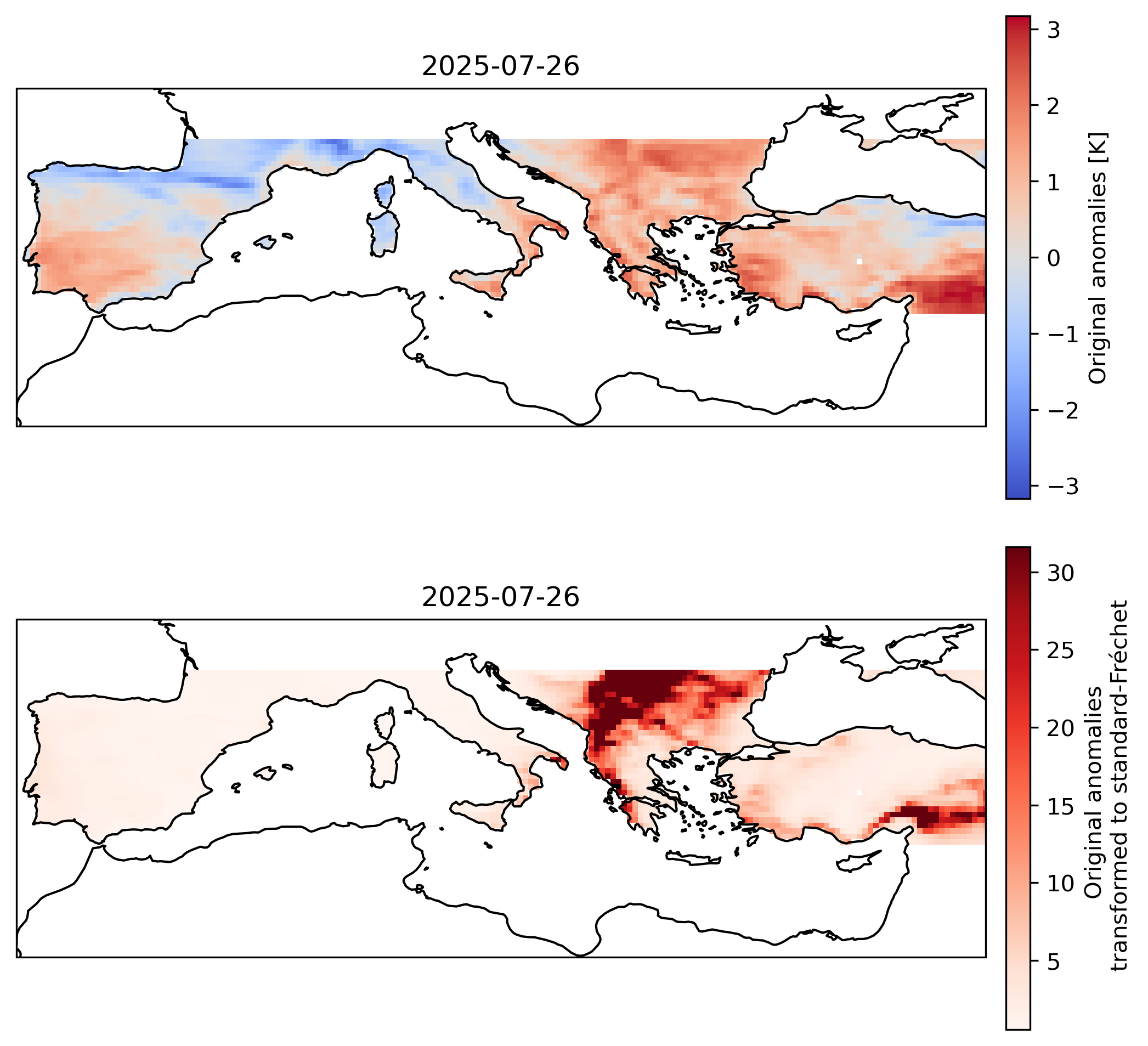}
    \caption{The effect of transforming temperature anomalies to a standard-Fr\'echet distribution using the ERA5 T2max on 26 July 2025.}
    \label{fig:effect_of_frechet}
\end{figure}

We consider a data set containing daily values. The annual cycle is initially removed from the daily data set during JJA. For this purpose, the T2max values for each day of the JJA period and grid point are standardised using the respective mean value and standard deviation. Subsequently, the data at each grid point are transformed to a standard distribution, since the estimator of the extremal dependence is based on the assumption that the margins are Fr\'echet distributed. In our case, this is a Fréchet distribution with tail index $2$. 
Transforming the data to Fr\'echet can be understood as giving little weight to small and negative temperature anomalies, and most weight to larger positive temperature anomalies. Figure~\ref{fig:effect_of_frechet} illustrates this effect.
 
The pairwise measure of dependence between extreme events at two locations is calculated for each pair of land grid points in Europe. The result is a matrix of extremal dependencies, referred to as the tail pairwise dependence matrix (TPDM), with dimensions $q \times q$, where $q$ is the number of grid points over land.
In the original paper that defined the principal component analysis (PCA) for extremes \citep{Cooley2019}, the same threshold was applied to all data pairs, which guarantees positive definiteness of the TPDM.
We use another estimator as advised by \citet{jiang_principal_2020}, who suggest using an individual threshold for each pair of data.
This estimator of the TPDM is not guaranteed to be positive definite; therefore, the positive definite matrix closest to the estimated one is used, which guarantees non-negative eigenvalues \citep[Sect.~2.2]{Szemkus2024}. The TPDM's eigenvalue decomposition then yields the eigenvectors and eigenvalues. The eigendecomposition procedure is similar to that of the standard PCA, since the TPDM satisfies all necessary properties, namely it is symmetric, has non-negative entries, and is positive semi-definite. The eigenvectors of the TPDM represent spatial variation patterns, indicating regions where extremes occur together with greater frequency.
 
The first mode has a non-zero spatial mean and values that remain nearly constant across space. The higher-order eigenvectors represent large-scale patterns associated with typical dipole and multipole structures, much like in standard PCA. In this way, we utilise the properties of the PCA, which favours large-scale patterns, to identify heat events that exhibit a correspondingly large spatial extent.
The eigenvalues $\lambda_k$, $k=1,\ldots,q$, indicate the proportion of pattern $k$ within the total variability of the standardised extreme events. The projection of the $k$-th pattern onto the Fr\'echet-standardised data yields the principal component (PC) $\eta_{k,i,d}$, i.e. the strength of the pattern $k$ on day $d$ of the season of year $i$.
Since the eigenvectors are
projected onto the Fr\'echet-standardised data rather than the original anomalies, the construction of the TPDM and the projection steps are based on the same data.
 
The TPDM eigenvectors span an `extreme' subspace that, based on our experience, can be adequately covered with $n_p=10$ extremal patterns. For example, these $n_p$ extremal patterns of ERA5 T2max in the southern European region account for about 75~\% of the total variance in the Fréchet-standardised data.
If the temperature field does not exhibit any large-scale positive temperature extremes, it is projected onto this subspace only to a limited extent. Small-scale extremes are intentionally not captured.
The total variability of the first $n_p$ extremal patterns is summarised in the extremal pattern index \citep[EPI,][]{Szemkus2024}.
The EPI on day $d$ of the summer season of year $i$ is defined as
\begin{equation}\label{eqn:EPI}
    \mathrm{EPI}_{i,d}
    =\frac{\sqrt{\sum_{k=1}^{n_p}\eta_{k,i,d}^{2}}}
          {\sqrt{\sum_{k=1}^{n_p}\lambda_{k}}} .
\end{equation}
The EPI represents a spatial aggregation of the extremal state of the variable in space \citep{Szemkus2024}.
Normalisation in the definition of the EPI using the sum of the eigenvalues in Eq.~(\ref{eqn:EPI}) allows the comparison of data on different spatial grids or of EPIs computed using different numbers of patterns.
The EPI allows multivariate extreme events with spatial extent to be described by a univariate index that has only a temporal dimension and accounts for spatial dependencies at extreme levels.
In this study, the method is applied to different AR6 regions (Fig.~\ref{fig:landmask}), i.e., the TPDM is calculated and decomposed separately for each region.
Throughout the remainder of this paper we write $y_{i,d}:=\mathrm{EPI}_{i,d}$; the EPI time series is thus the quantity to which the extreme value model of Sect.~\ref{sec:theory} is applied.
 
\subsubsection{Example: Summer 2025 in Europe}
\label{sec:example2025}
\begin{figure*}[t]
    \includegraphics[width=1\textwidth]{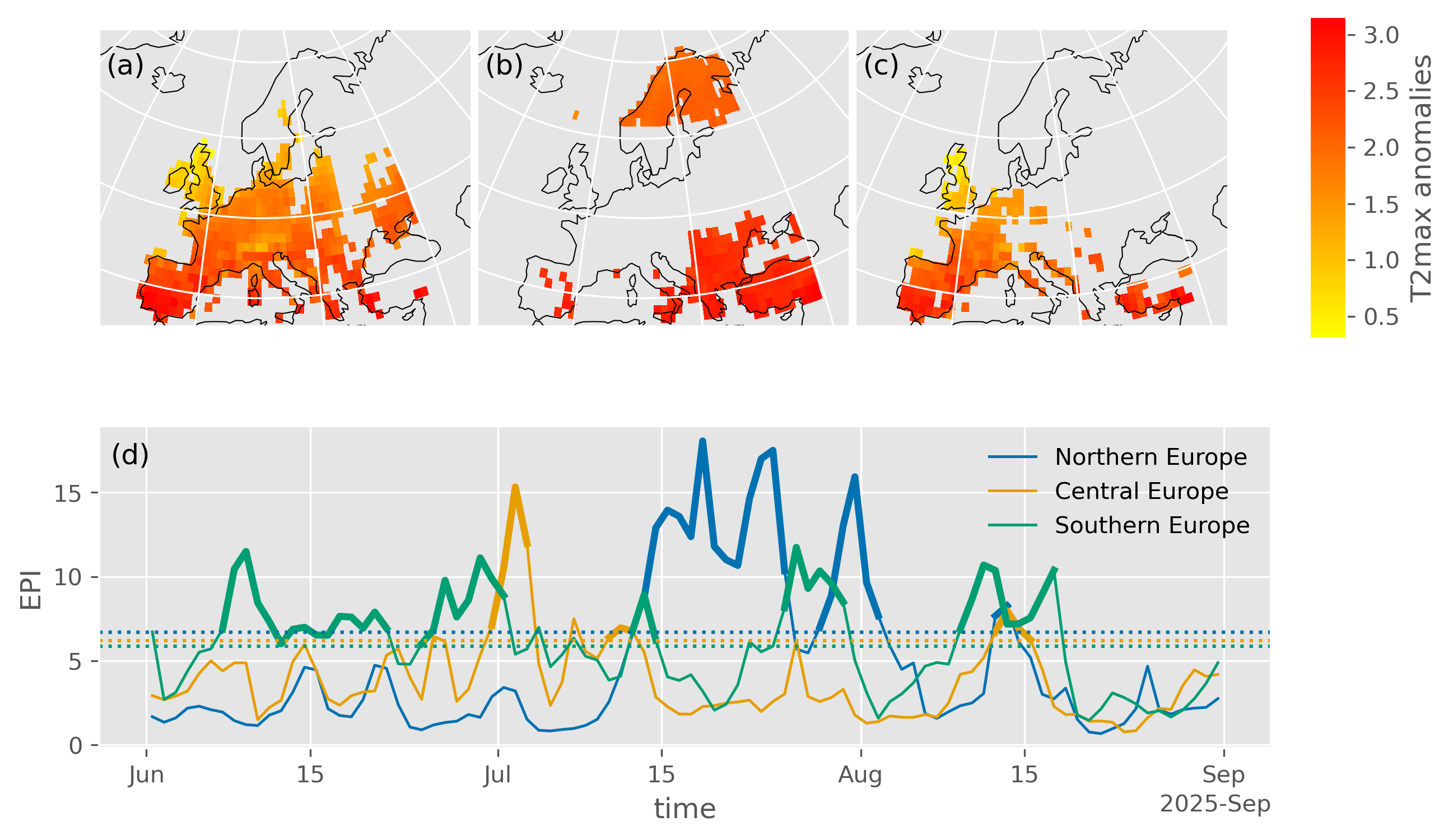}
    \caption{Mean standardised T2max anomalies in ERA5 between (a) 7 June to 5 July 2025, (b) 15 July to 1 August 2025, and (c) 7 August to 19 August 2025. Only grid points exceeding the 99~\% quantile are shown. In (d) the EPI for T2max from June to August 2025 for northern (blue), central (orange) and southern (green) European regions from Fig.~\ref{fig:landmask} is shown. The EPI is bold if it exceeds the corresponding 95~\% quantile (1940--2025) of the region, with the corresponding quantile indicated by a dotted line.}\label{fig:anomalies}
\end{figure*}

The European summer of 2025 was characterised by a series of heatwave events across Europe. An early heatwave began in southwestern Europe in June, and the highest average surface air temperatures between 17 June and 2 July 2025 compared to the period 1979--2024 were recorded on the Iberian Peninsula and in large parts of France, extending to Germany and southern Great Britain, as reported in \citet{Copernicus2025Heatwaves}.
Figure \ref{fig:anomalies} shows the EPI for the regions in Fig.~\ref{fig:landmask} and the T2max anomalies for summer 2025 over Europe.
The early onset of the remarkable heatwave over southwestern Europe is visible at the end of the first week of June (Fig.~\ref{fig:anomalies}). This event lasted until the beginning of July. As is common in this region, this heatwave was mainly caused by atmospheric ridges, as is evident from the large-scale circulation pattern during the heatwave \citep{faranda:hal-05289765}. Regarding the spatial extent, most parts of the southern and central European regions except Turkey were influenced by this event.
Two additional strong heatwave events were observed in southern Europe, one in the second half of July and one in the first half of August. The event in July occurred over the eastern Mediterranean, while the event in August took place over the western Mediterranean, affecting Spain, Portugal, and France.
 
Another remarkable heatwave in 2025 occurred over Scandinavia, from approximately mid-July to early August. This event received considerable media attention and was investigated by the World Weather Attribution in a report \citep{report2025}. A detailed description of the synoptic situation and the resulting temperature extremes can be found in the report.
A detailed scientific analysis of the European summer of 2025, including the characterisation of heat events, their impact, and the influence of climate warming, can be found in the ClimXtreme report by \citet{Lemburg2026}.

\section{Theory and Methods}\label{sec:theory}

\subsection{Attribution of extreme events}
\label{sec:theory_attribution}
 
The attribution of extreme events corresponds to the statistical answer to the question of whether or not a specific event was caused by anthropogenic climate change. The theoretical foundation of the attribution is presented in detail by \citet{ProbabilitiesofCausationofClimateChanges} and is based on causal counterfactual theory.
 
We define $m_0$ as the scenario without anthropogenic forcing (HIST-NAT) and $m_1$ as the scenario with climate change (HIST). The probability ratio is then the ratio of the probability of an event under HIST to the probability of an event under HIST-NAT, given by
\begin{equation}
    \mathrm{PR} = \frac{p_1}{p_0}.
\end{equation}
 
In this study, we model time series rather than single events. We therefore consider the joint density of a set of time series $\boldsymbol{Y}=(\boldsymbol{y}^{(1)} , \ldots , \boldsymbol{y}^{(n_y)})$ under the two scenarios HIST and HIST-NAT, where the vector $\boldsymbol{y}^{(i)}$ denotes the daily evolution of the EPI during the season of year $i$. The probability ratio then becomes a likelihood ratio (LR):
\begin{equation}\label{eqn:probratio}
    \mathrm{LR}(\boldsymbol{Y})=\frac{l(\boldsymbol{Y}\mid m_1)}{l(\boldsymbol{Y}\mid m_0)},
\end{equation}
where $l(\boldsymbol{Y}\mid m_s)$ is the likelihood of observing $\boldsymbol{Y}$ under scenario $m_s$, $s\in\{0,1\}$.
 
From a Bayesian perspective \citep{min2004, Seong2022}, attribution can instead be stated as the ratio of the posterior odds for the scenarios \citep{kass_bayes_1995}. Here, we treat $m_0$ and $m_1$ as two competing scenario-conditioned generative statistical models for the observed ERA5 EPI time series. We assume that $(\boldsymbol{y}^{(1)} , \ldots , \boldsymbol{y}^{(n_y)})$ are independent vectors: the daily time series of one season is independent of the daily time series of the same season in the year before or after. Using Bayes' theorem, the posterior odds can be written as
\begin{equation}\label{eqn:BayesII}
 \frac{\mathbb{P}(m_1\mid\boldsymbol{Y})}{\mathbb{P}(m_0\mid\boldsymbol{Y})}
 =\frac{l(\boldsymbol{Y}\mid m_1)}{l(\boldsymbol{Y}\mid m_0)}\cdot
  \frac{\mathbb{P}(m_1)}{\mathbb{P}(m_0)}
 =\prod_{i=1}^{n_y}\frac{l(\boldsymbol{y}^{(i)}\mid m_1)}{l(\boldsymbol{y}^{(i)}\mid m_0)}\cdot
  \frac{\mathbb{P}(m_1)}{\mathbb{P}(m_0)}
 =\underbrace{\prod_{i=1}^{n_y}\mathrm{LR}\bigl(\boldsymbol{y}^{(i)}\bigr)}_{=:\ \mathrm{BF}}
  \cdot\frac{\mathbb{P}(m_1)}{\mathbb{P}(m_0)} .
\end{equation}
 
The likelihood ratio $\mathrm{LR}(\boldsymbol{y}^{(i)}) = l(\boldsymbol{y}^{(i)}\mid m_1)/l(\boldsymbol{y}^{(i)}\mid m_0)$ represents the additional evidence contributed by the seasonal time series of the year $i$ to the total evidence. Under uniform prior probabilities, the posterior ratio equals the product of the likelihood ratios, which corresponds to the Bayes factor (BF).
By varying the most recent analysis year from 1940 to 2025, we can assess when in the past under today's knowledge would have provided sufficient evidence of climate change.
 
A statement of evidence based on the Bayes factor is given in Table~\ref{tab:bayesfactor}, which differs slightly from that given in \citet{kass_bayes_1995}. Given that interpretation is context-dependent, Table~\ref{tab:bayesfactor} provides a general indication of the level of evidence.
\begin{table}[h]
\caption[Scales of the Bayes factor and related evidence.]{Scales of the Bayes factor and related evidence as given in {\citet{Seong2022}}. Here and throughout, $\log$ denotes the natural logarithm.}\label{tab:bayesfactor}
\small
\centering
\begin{tabular}{ |l||l||l|  }
 \hline
    $\log\mathrm{BF}$ & $\mathrm{BF}$ & Evidence against $m_0$\\
     \hline
    0--1  & 1--3 & Not worth more than a bare mention \\
    1--2.5  & 3--12 & Substantial \\
    2.5--5  & 12--150 & Strong \\
    $>5$  & $>150$ & Decisive \\
    \hline
    \end{tabular}
\end{table}

\subsection{Likelihood formulation}\label{sec:Markov}
 
\subsubsection{Marginal formulation}
\label{sec:marginal_formulation}

Our aim is to model the extremes of the EPI time series, as illustrated in Fig.~\ref{fig:anomalies} by the bold sections of the respective time series that exceed the corresponding threshold.
When modelling extremes over a high threshold, we use a common peaks-over-threshold (POT) approach, in which threshold exceedances follow a generalized Pareto distribution (GPD). 

We write the full data set as
\begin{equation}\label{eqn:Ydef}
  \boldsymbol{Y}=\{y_{i,d}\,;\ i=1,\ldots,n_y,\ d=1,\ldots,n\},
\end{equation}
where $y_{i,d}$ is the EPI on day $d$ of the summer season of year $i$, $n_y$ is the number of JJA
seasons and $n=92$ the number of days per season. The time series of a single season is
$\boldsymbol{y}^{(i)}=(y_{i,1},\ldots,y_{i,n})$, so that
$\boldsymbol{Y}=(\boldsymbol{y}^{(1)},\ldots,\boldsymbol{y}^{(n_y)})$ as in
Sect.~\ref{sec:theory_attribution}. Season $i=1$ corresponds to the summer of 1940 and season
$i=n_y$ to the summer of 2025. An overview of the notation used throughout this paper is
provided in Tables~\ref{tab:notation_data} and~\ref{tab:notation_params} in
Appendix~\ref{sec:notation_overview}.
 
Let $u_{i,d}$ denote a sufficiently high threshold for observation $d$ in year $i$. Conditional
on exceeding $u_{i,d}$, the excess $y_{i,d}-u_{i,d}$ is modelled by a GPD with the scale parameter
$\sigma_{i,d}>0$ and the shape parameter $\xi_{i,d}$. Together with the exceedance probability
$\phi_{i,d}=\mathbb{P}(y_{i,d}>u_{i,d})$, the marginal distribution function reads
\citep[Sect.~4.3]{Coles2001}
\begin{equation}\label{eqn:marginal_gpd}
F(y_{i,d})\approx
\begin{cases}
1-\phi_{i,d}, & y_{i,d}\le u_{i,d},\\[0.8em]
1-\phi_{i,d}\left(1+\xi_{i,d}\dfrac{y_{i,d}-u_{i,d}}{\sigma_{i,d}}\right)^{-1/\xi_{i,d}}_{+},
& y_{i,d}>u_{i,d}.
\end{cases}
\end{equation}
Thus $\phi_{i,d}$ is the probability that observation $d$ in year $i$ exceeds the threshold
$u_{i,d}$; below the threshold the distribution is treated in censored form, so that only the
probability of non-exceedance enters the model. Here $F$ always denotes a univariate marginal
distribution function; the joint distribution of two consecutive observations is denoted by $G$
in Sect.~\ref{sec:dependence_formulation}. At this stage, all four quantities are allowed to
depend on both the year and the day; the assumptions that reduce this dependence are stated
explicitly in Sect.~\ref{sec:nonstat}.

\subsubsection{Dependence formulation}
\label{sec:dependence_formulation}
In practice, extreme heat often persists for several consecutive days.
Consequently, several threshold exceedances may occur in sequence and cannot be treated as independent events. Since the temporal dependence is inherent in the data, we aim to model it explicitly, as the duration of a heatwave is a key factor in determining its impact. 
For alternative approaches to modelling the temporal dependence of a time series, we refer the reader to \citet{Beirlant}. 

We model the temporal dependence of the threshold exceedances as a first-order Markov process:
the seasons are independent across the year index $i$, while within a season the values
$y_{i,1},\ldots,y_{i,n}$ form a Markov process. The first-order Markov property implies that the conditional
density of day $d$ given all previous days reduces to the density conditioned on the previous
day only, $l(y_{i,d}\mid y_{i,1},\ldots,y_{i,d-1};\boldsymbol{\theta})
=l(y_{i,d}\mid y_{i,d-1};\boldsymbol{\theta})$. Writing $l_1$ and $l_2$ for the univariate and
bivariate densities implied by the model, the likelihood of the season $i$ is
\begin{equation}\label{eqn:markov_chain}
\begin{split}
l\bigl(\boldsymbol{y}^{(i)};\boldsymbol{\theta}\bigr)
&= l_1(y_{i,1};\boldsymbol{\theta}_1)\prod_{d=2}^{n}
   l\bigl(y_{i,d}\mid y_{i,d-1};\boldsymbol{\theta}\bigr)
 = l_1(y_{i,1};\boldsymbol{\theta}_1)\prod_{d=2}^{n}
   \frac{l_2(y_{i,d-1},y_{i,d};\boldsymbol{\theta})}{l_1(y_{i,d-1};\boldsymbol{\theta}_1)}\\
&= \frac{\prod_{d=1}^{n-1} l_2(y_{i,d},y_{i,d+1};\boldsymbol{\theta})}
        {\prod_{d=2}^{n-1} l_1(y_{i,d};\boldsymbol{\theta}_1)},
\end{split}
\end{equation}
and the likelihood of the complete data set is
\begin{equation}\label{eqn:full_likelihood}
l(\boldsymbol{Y};\boldsymbol{\theta})=\prod_{i=1}^{n_y} l\bigl(\boldsymbol{y}^{(i)};\boldsymbol{\theta}\bigr).
\end{equation}
Here $\boldsymbol{\theta}=(\boldsymbol{\theta}_1,\boldsymbol{\theta}_2)$, with
$\boldsymbol{\theta}_1$ the marginal parameters of Sect.~\ref{sec:marginal_formulation} and
$\boldsymbol{\theta}_2$ the possibly infinite set of parameters of the temporal extremal dependence. Due to the
first-order assumption, the dependence is fully described by the joint distribution of pairs of
consecutive days $(y_{i,d},y_{i,d+1})$.

We use bivariate extreme value theory for the dependence formulation.
Unlike the univariate case, multivariate extreme value theory does not provide a single, complete parametric model for the dependence structure.
Instead, extremal dependence can be described either non-parametrically or by assuming a parametric family for the stable tail dependence function.
In this study, we focus on a parametric model of temporal extremal dependence within the class of asymptotically dependent bivariate extreme-value models.
This assumption is consistent with the pre-asymptotic nature of the data under consideration, in which exceedances occur above a high but finite threshold.
 
Consider two consecutive observations $(y_{i,d},y_{i,d+1})$ with a joint distribution function $G$, and assume that $G$ lies in the domain of attraction of a bivariate extreme value distribution. Before specifying the dependence structure, the margins are transformed to standard Fr\'echet.
 
The transformation to standard Fr\'echet is directly linked to the GPD formulation in Eq.~(\ref{eqn:marginal_gpd}), given by
\begin{equation}\label{eqn:margin}
v_{i,d}=\phi_{i,d}\left(1+\xi_{i,d}\frac{y_{i,d}-u_{i,d}}{\sigma_{i,d}}\right)^{-1/\xi_{i,d}}_{+},
\end{equation}
so that $F(y_{i,d})\approx 1-v_{i,d}$ and $1/v_{i,d}$ is standard Fr\'echet distributed
\citep[Sect.~10.4]{Beirlant}. Note that $v_{i,d}=\phi_{i,d}$ whenever $y_{i,d}=u_{i,d}$, which
will be used in Eq.~(\ref{eqn:markov}) below.
This separate modelling of the marginal distribution and extremal dependence corresponds to the approach used in deriving the TPDM in Sect.~\ref{sec:EPI}.

The joint distribution of two consecutive observations above high thresholds is then
approximated by
\begin{equation}\label{eqn:full_tail}
G(y_{i,d},y_{i,d+1})\approx\exp\bigl\{-V(v_{i,d},v_{i,d+1})\bigr\},
\end{equation}
where $V$ denotes the stable tail dependence function. Since the marginal behaviour has already been absorbed into the transformation to $v_{i,d}$ and $v_{i,d+1}$, the function $V$ describes the extremal dependence between two consecutive observations. Its significance lies in the fact that a bivariate extreme value distribution is completely determined by its margins and $V$. For the function $V$ to be a stable tail dependence function in Eq.~(\ref{eqn:full_tail}), it must fulfil certain necessary conditions. The stable tail dependence function restricted to the unit simplex is also 
known as the more popular Pickands dependence function, originally introduced by \citet{pickands1981multivariate}, see \citet[Sect.~8.2.5]{Beirlant} for details. 

In practice, the function $V$ can be expressed in non-parametric or parametric form. 
Our focus is on parametric families of extremal dependence, which can be conveniently specified through $V$. The logistic model introduced in the following is an example of this. 
The logistic model as the simplest parametric choice of the function $V$ is given as 
\citep[Sect.~10.4.5]{Beirlant},
\begin{equation}\label{eqn:logistic_tail}
V(v_{i,d},v_{i,d+1})
=\left(v_{i,d}^{1/\alpha_{i,d}}+v_{i,d+1}^{1/\alpha_{i,d}}\right)^{\alpha_{i,d}},
\qquad v_{i,d},v_{i,d+1}\ge 0,
\end{equation}
where $\alpha_{i,d}\in(0,1]$ is the single dependence parameter attached to the transition from
day $d$ to day $d+1$ in year $i$. The case $\alpha_{i,d}=1$ corresponds to independence, whereas
$\alpha_{i,d}\rightarrow 0$ corresponds to complete asymptotic dependence. With
Eqs.~(\ref{eqn:margin}) and (\ref{eqn:logistic_tail}), the bivariate model in
Eq.~(\ref{eqn:full_tail}) is fully determined.
 
More flexible parametric choices could also be used, such as asymmetric logistic or negative logistic models, as well as non-parametric descriptions of the stable tail dependence function.
For alternative choices of $V$, the reader is referred to \citet[Sect.~9.2]{Beirlant}.
In this study, the logistic model is used because it provides the most straightforward parametric description of the temporal extremal dependence, requiring only one dependence parameter.
 
Max-stable models, such as the logistic model, cannot capture asymptotic independence \citep[Sect.~2.3]{https://doi.org/10.1111/sjos.12491}.
They can represent either complete independence or asymptotic dependence.
Recent studies emphasise the importance of models that can represent both asymptotic dependence and independence \citep[e.g.,][]{ledford_tawn2,bortor_tawn,ramos_ledford,Ramos2013,Wadsworth}.
We will not use this extension here, but we consider it an interesting avenue for future studies. 
 
\subsubsection{Likelihood inference}
\label{sec:inference}
\begin{figure}
    \centering
    \includegraphics[width=0.76\linewidth]{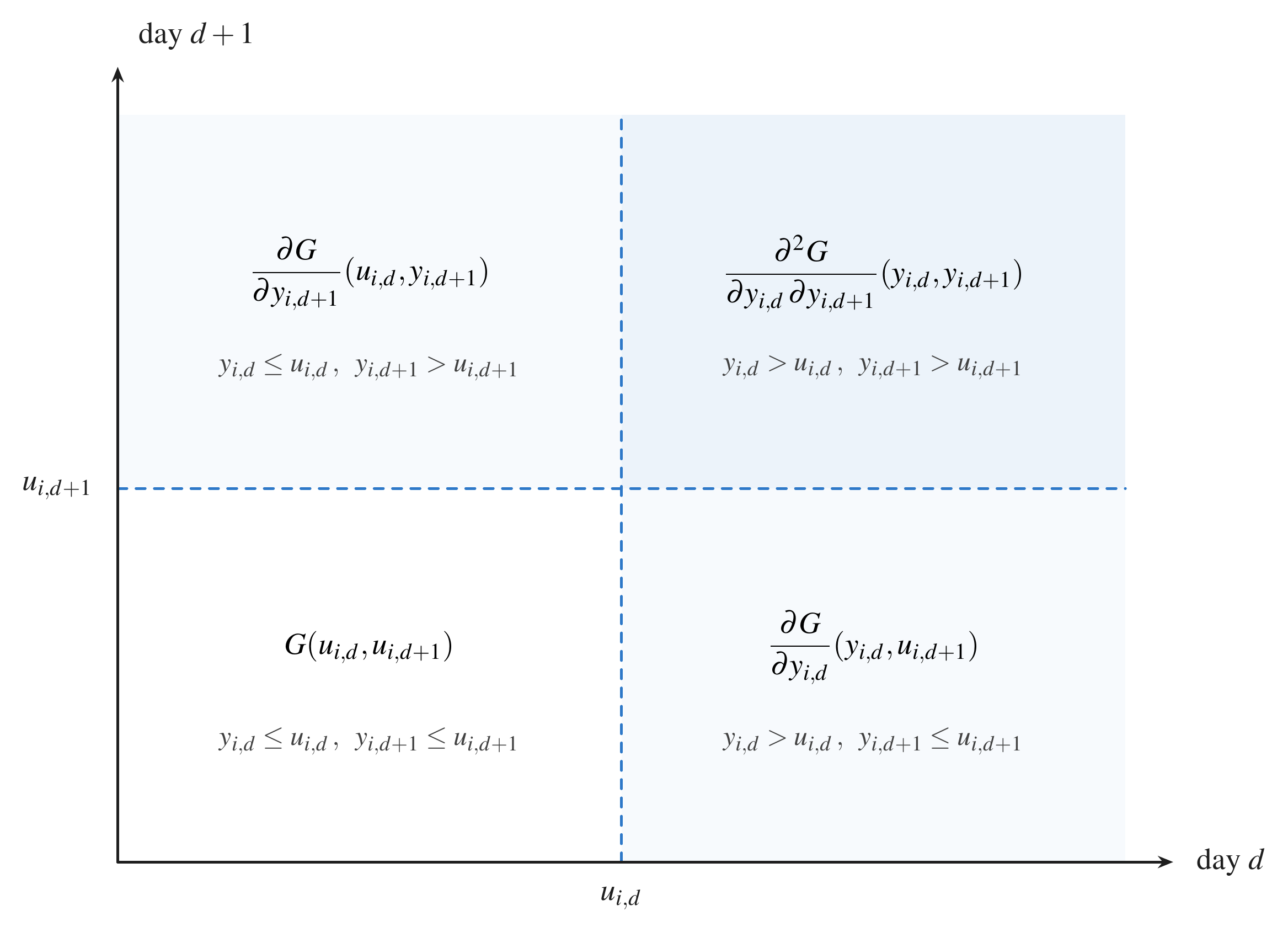}
    \caption{Regions of the censored likelihood model according to Eq.~(\ref{eqn:contributions}), where the likelihood contribution depends on which of the two variables exceeds its threshold.}
    \label{fig:censored_likelihood_diagram}
\end{figure}

For the likelihood inference, we focus on the censored likelihood method.
This model can handle all situations: when neither of the two consecutive days exceeds the threshold; when one of them exceeds the threshold; and when both exceed the threshold.
An advantage of the censoring is that non-extreme observations do not affect the estimation of the extremal dependence.
Depending on which of the two consecutive observations exceeds its threshold, the bivariate plane is divided into the four regions shown in Fig.~\ref{fig:censored_likelihood_diagram}, with the corresponding likelihood contributions
\begin{equation}\label{eqn:contributions}
l_2(y_{i,d},y_{i,d+1};\boldsymbol{\theta})\propto
\begin{cases}
G(u_{i,d},u_{i,d+1}) & \text{if } y_{i,d}\le u_{i,d},\ y_{i,d+1}\le u_{i,d+1},\\[0.4em]
\dfrac{\partial G}{\partial y_{i,d}}(y_{i,d},u_{i,d+1})
 & \text{if } y_{i,d}>u_{i,d},\ y_{i,d+1}\le u_{i,d+1},\\[0.6em]
\dfrac{\partial G}{\partial y_{i,d+1}}(u_{i,d},y_{i,d+1})
 & \text{if } y_{i,d}\le u_{i,d},\ y_{i,d+1}>u_{i,d+1},\\[0.6em]
\dfrac{\partial^2 G}{\partial y_{i,d}\,\partial y_{i,d+1}}(y_{i,d},y_{i,d+1})
 & \text{if } y_{i,d}>u_{i,d},\ y_{i,d+1}>u_{i,d+1}.
\end{cases}
\end{equation}
 
Writing $V_1$, $V_2$ and $V_{12}$ for the partial derivatives of $V$ with respect to its first, second and both arguments, the contributions follow from Eqs.~(\ref{eqn:full_tail}) and (\ref{eqn:contributions}) as \citep{Beirlant}
\begin{equation}\label{eqn:markov}
l_2(y_{i,d},y_{i,d+1};\boldsymbol{\theta})\propto
\begin{cases}
\exp\{-V(\phi_{i,d},\phi_{i,d+1})\}
 & \text{if } y_{i,d}\le u_{i,d},\ y_{i,d+1}\le u_{i,d+1},\\[0.6em]
-\dfrac{\partial v_{i,d}}{\partial y_{i,d}}\,V_1(v_{i,d},\phi_{i,d+1})\,
 \exp\{-V(v_{i,d},\phi_{i,d+1})\}
 & \text{if } y_{i,d}>u_{i,d},\ y_{i,d+1}\le u_{i,d+1},\\[0.9em]
-\dfrac{\partial v_{i,d+1}}{\partial y_{i,d+1}}\,V_2(\phi_{i,d},v_{i,d+1})\,
 \exp\{-V(\phi_{i,d},v_{i,d+1})\}
 & \text{if } y_{i,d}\le u_{i,d},\ y_{i,d+1}>u_{i,d+1},\\[0.9em]
\dfrac{\partial v_{i,d}}{\partial y_{i,d}}\dfrac{\partial v_{i,d+1}}{\partial y_{i,d+1}}
 \bigl[V_1V_2-V_{12}\bigr](v_{i,d},v_{i,d+1})\,\exp\{-V(v_{i,d},v_{i,d+1})\}
 & \text{if } y_{i,d}>u_{i,d},\ y_{i,d+1}>u_{i,d+1},
\end{cases}
\end{equation}
where the first case uses $v_{i,d}=\phi_{i,d}$ when $y_{i,d}=u_{i,d}$. 
 
Thus, every pair of consecutive time series points $l_2(y_{i,d}, y_{i,d+1};\boldsymbol{\theta})$ must be evaluated.
For the univariate likelihood $l_1(y_{i,d};\boldsymbol{\theta}_1)$, we can write
\begin{equation}\label{eqn:univ_density}
l_1(y_{i,d};\boldsymbol{\theta}_1)=
\begin{cases}
\dfrac{\phi_{i,d}}{\sigma_{i,d}}
 \left(1+\xi_{i,d}\dfrac{y_{i,d}-u_{i,d}}{\sigma_{i,d}}\right)_{+}^{-1/\xi_{i,d}-1},
 & \text{if } y_{i,d}>u_{i,d},\\[1em]
1-\phi_{i,d}, & \text{if } y_{i,d}\le u_{i,d}.
\end{cases}
\end{equation}

The full likelihood used for the fitting procedure is given by
Eqs.~(\ref{eqn:markov_chain}) and (\ref{eqn:full_likelihood}), with the bivariate contributions
of Eq.~(\ref{eqn:markov}) in the numerator and the univariate contributions of
Eq.~(\ref{eqn:univ_density}) in the denominator. Since the Markov process conditions on the
previous day, the product of pairwise bivariate contributions has to be divided by the
univariate ones. Because the bivariate density is required in the numerator, the distribution
function must be differentiated with respect to those variables that exceed the threshold,
following the censoring scheme of Fig.~\ref{fig:censored_likelihood_diagram}. The marginal and
dependence parameters are finally obtained from
\begin{equation}\label{eqn:optimisation}
    \hat{\boldsymbol{\theta}}
    =\arg\min_{\boldsymbol{\theta}}\bigl\{-\log l(\boldsymbol{Y};\boldsymbol{\theta})\bigr\}.
\end{equation}

\subsubsection{Higher-order Markov processes}
\label{sec:higher_order}
The first-order assumption of the Markov process may be questionable. To investigate the effect of this choice, we expand the analysis to include a second-order Markov process to capture higher-order dependencies on a longer timescale. To this end, we can rewrite Eq.~(\ref{eqn:markov_chain}) to
\begin{equation}\label{eqn:markov_chain_2}
\begin{split}
l\bigl(\boldsymbol{y}^{(i)};\boldsymbol{\theta}\bigr)
&= l_2(y_{i,1},y_{i,2};\boldsymbol{\theta})\prod_{d=3}^{n}
   l\bigl(y_{i,d}\mid y_{i,d-1},y_{i,d-2};\boldsymbol{\theta}\bigr)
 = l_2(y_{i,1},y_{i,2};\boldsymbol{\theta})\prod_{d=3}^{n}
   \frac{l_3(y_{i,d-2},y_{i,d-1},y_{i,d};\boldsymbol{\theta})}
        {l_2(y_{i,d-2},y_{i,d-1};\boldsymbol{\theta})}\\
&= \frac{\prod_{d=1}^{n-2} l_3(y_{i,d},y_{i,d+1},y_{i,d+2};\boldsymbol{\theta})}
        {\prod_{d=2}^{n-2} l_2(y_{i,d},y_{i,d+1};\boldsymbol{\theta})},
\end{split}
\end{equation}
with $l_3$ the trivariate density. The likelihood ratio can then be calculated using this
likelihood formulation.
 
As described in \citet{Smith_markov}, the estimation of the model is similar to that for a first-order Markov process, but now also with respect to the newly introduced variable (trivariate problem).
In this case, the construction of Fig.~\ref{fig:censored_likelihood_diagram}, which is not unique to the
two-dimensional case, can be extended to more than two variables.

For the symmetric logistic model, an example of a corresponding tail dependence function is given by
\begin{equation}\label{eqn:logistic_tail_3}
V(v_{i,d},v_{i,d+1},v_{i,d+2})
=\left(v_{i,d}^{1/\alpha_{i,d}}+v_{i,d+1}^{1/\alpha_{i,d}}+v_{i,d+2}^{1/\alpha_{i,d}}\right)^{\alpha_{i,d}},
\qquad v_{i,d},v_{i,d+1},v_{i,d+2}\ge 0 .
\end{equation}
The symmetric logistic model imposes the same dependence on all pairs among three consecutive days, although the dependence between days $d$ and $d+2$ would be expected to be weaker than between adjacent days.
 
However, due to the more complex likelihood structure, the fitting procedure is more costly than for a first-order Markov process, and the likelihoods of different orders are not necessarily comparable \citep[Sect.~10.4.6]{Beirlant}.
The results for higher-order Markov processes are presented in Sect.~\ref{sec:second_order}.

\subsection{Including non-stationarity}
\label{sec:nonstat}
 
In the general formulation of Sect.~\ref{sec:Markov} every parameter carries both indices,
i.e. $u_{i,d}$, $\phi_{i,d}$, $\sigma_{i,d}$, $\xi_{i,d}$ and $\alpha_{i,d}$. In the main part
of this study we assume that the parameters vary between seasons but are constant within a
season,
\begin{equation}\label{eqn:no_seasonal}
u_{i,d}\equiv u_i,\quad \phi_{i,d}\equiv\phi_i,\quad \sigma_{i,d}\equiv\sigma_i,\quad
\xi_{i,d}\equiv\xi_i,\quad \alpha_{i,d}\equiv\alpha_i ,
\end{equation}
so that the day index $d$ can be dropped. The consequences of relaxing assumption
(\ref{eqn:no_seasonal}), i.e. of allowing for a residual seasonal cycle, are examined in
Appendix~\ref{sec:seasonal_cycle}.
 
To account for non-stationarity due to the general warming trend, we include slow changes over time in the marginal and dependence parameters. Since the exceedance-probability and the threshold are time-dependent, one of the two is specified a priori. We
investigate two modelling approaches:
\begin{enumerate}
    \item $u=\text{const}$, $\phi_i$ time-varying: the threshold is fixed a priori as the
    95~\% quantile, and the exceedance probability is estimated as a time-varying parameter using logistic regression
    (Sect.~\ref{sec:using_constant_threshold}).
    \item $\phi=\text{const}$, $u_i$ time-varying: the exceedance probability is fixed a priori
    at $\phi=0.05$, and the warming trend is absorbed into a time-varying threshold $u_i$
    estimated by quantile regression (Sect.~\ref{sec:using_time_varying_threshold}).
\end{enumerate}

\begin{figure}[t]
    \centering
    \includegraphics[width=0.8\linewidth]{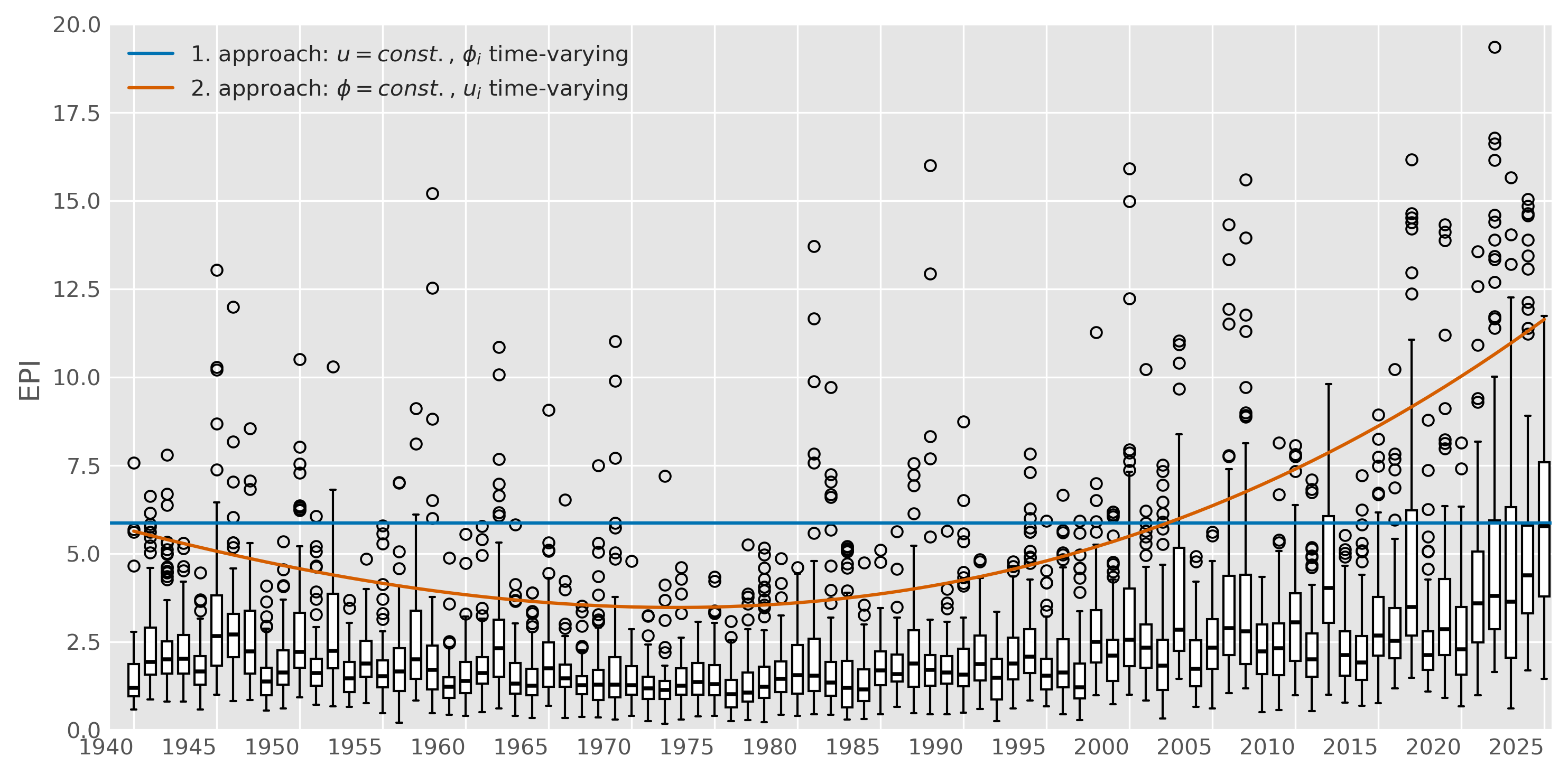}
    \caption{The ERA5 EPI for the southern European region is plotted as a box plot for each summer. In addition, the two thresholds from the two modelling approaches are shown. The first model's constant threshold is shown in blue and the second model's time-varying threshold is shown in orange, both for the 95~\% quantile.}
    \label{fig:comparison_approaches}
\end{figure}

The two modelling approaches are visualised by the constant and time-varying thresholds in Fig.~\ref{fig:comparison_approaches}, which is based on the ERA5 EPI in the southern European region. Both modelling approaches become visible. When the threshold is constant (blue line), the number of threshold exceedances has increased since 2000. When the threshold is time-varying (orange line), the number of threshold exceedances remains nearly constant. The slowly varying threshold shows the cooler decades between 1960 and 1990, as well as the warming trend since the 1990s.
The reasoning behind the non-stationary modelling of exceedance probability and threshold can be found in Sect.~\ref{sec:reasoning}. In this context, we refer to \citet{Eastoe}, who discuss the two different modelling approaches in
more detail. 
The time-varying parameters are modelled through slowly varying covariates. We use low-order
Legendre polynomials \citep{Min2006}, which are well suited to model non-linear trends because
they are orthogonal, data-independent, and smooth. Let
\begin{equation}\label{eqn:legendre}
x_i=\frac{2(i-1)}{n_y-1}-1\in[-1,1],\qquad i=1,\ldots,n_y,
\end{equation}
denote the season index rescaled to $[-1,1]$, the interval on which the Legendre polynomials
are orthogonal, and let
$\boldsymbol{g}_i=\bigl(g_0(x_i),\ldots,g_K(x_i)\bigr)^{T}$ collect the polynomials of degree
$0$ to $K$ for the year $i$. We set $K=5$ throughout the article.

\subsubsection{Can extremes be attributed, and is their frequency increasing over time? -- Using a constant threshold}
\label{sec:using_constant_threshold}
With a constant
threshold $u$ in a non-stationary climate, the probability of exceeding this threshold
$\phi_{i} = \mathbb{P}(y_{i,d} > u)$, which under assumption (\ref{eqn:no_seasonal}) is the
same for all days $d$ of season $i$, varies over time and can therefore
no longer be treated as a fixed model parameter.
Since we assume that the exceedance probability is affected by the general warming trend, we model the non-stationarity using logistic regression with Legendre polynomials as covariates.
Logistic regression is a special case of a generalised linear model \citep{Nelder1972}. We define the logistic regression using a logit link function such that
\begin{equation}\label{eqn:logit}
    \mathrm{logit}(\phi_{i})
    =\log\!\left(\frac{\phi_{i}}{1-\phi_{i}}\right)
    =\boldsymbol{\beta}_\phi^T \boldsymbol{g}_i ,
    \qquad i=1,\ldots,n_y,
\end{equation}
with $\boldsymbol{\beta}_\phi=(\beta_{\phi,0},\ldots,\beta_{\phi,K})^T$. The coefficients
$\boldsymbol{\beta}_\phi$ are jointly estimated from the daily binary exceedance indicators
$\mathbf{1}\{y_{i,d}>u\}$ of all seasons. The scale and shape
parameters are modelled as
\begin{equation}\label{eqn:sigma_xi}
\sigma_{i}=\exp\bigl(\boldsymbol{\beta}_{\sigma}^T\boldsymbol{g}_i\bigr),
\qquad
\xi_{i}=\boldsymbol{\beta}_{\xi}^T\boldsymbol{g}_i ,
\qquad i=1,\ldots,n_y,
\end{equation}
with $\boldsymbol{\beta}_{\sigma} = (\beta_{\sigma,0}, \ldots, \beta_{\sigma,K})^T$ and
$\boldsymbol{\beta}_{\xi} = (\beta_{\xi,0}, \ldots, \beta_{\xi,K})^T$ as the vectors of regression
coefficients.
 
The dependence parameter $\alpha_i$ is also assumed to vary over time, assuming that the dependence structure between two consecutive days can change slowly over time.
Using a sigmoid link to ensure $\alpha_i\in(0,1)$, we can write
\begin{equation}\label{eqn:alpha}
    \alpha_{i}=\bigl[1+\exp\bigl(-\boldsymbol{\beta}_{\alpha}^T\boldsymbol{g}_i\bigr)\bigr]^{-1},
    \qquad i=1,\ldots,n_y.
\end{equation}
Note that the sigmoid link restricts $\alpha_i$ to the open interval $(0,1)$; exact independence ($\alpha_i=1$, cf. Eq.~\ref{eqn:logistic_tail}) is thus attained only in the limit $\boldsymbol{\beta}_{\alpha}^{T}\boldsymbol{g}_i\rightarrow\infty$.
The parameter vectors are therefore
$\boldsymbol{\theta}_1=(\boldsymbol{\beta}_\phi,\boldsymbol{\beta}_{\sigma},\boldsymbol{\beta}_{\xi})$
and $\boldsymbol{\theta}_2=\boldsymbol{\beta}_{\alpha}$. 
 
\subsubsection{Is there a climate change signal in the tail behaviour of heatwave extremes beyond a general shift in the temperature distribution? -- Using a variable threshold}
\label{sec:using_time_varying_threshold}
Since large parts of the attribution statement are due to changes in the probability of
exceeding the threshold $u$, we would like to exclude this effect and define the extremes on the
original scale \citep{Eastoe}. The attribution question that we ask in this case is as follows. Is there a
change in the tail behaviour of heatwave extremes beyond a general shift in the temperature
distribution?
To model only distributional changes above a time-varying threshold $u_i$, we define $u_i$ such
that the non-exceedance probability $\mathbb{P}(y_{i,d}\le u_i)=\tau$, which under assumption
(\ref{eqn:no_seasonal}) is the same for all days $d$ of season $i$, is constant, and hence
$\phi=1-\tau$. We obtain $u_i$ as the conditional $\tau$-quantile by quantile regression
\citep{Koenker1999,Koenker2005}, following \citet[Sect.~7.4.2]{Beirlant}. As
\citet{Chavez-Demoulin_Davison_2012} point out, a time-dependent threshold is preferable for a
more precise estimation of regression effects.
 
Quantile regression assumes a linear model for the conditional quantile,
\begin{equation}\label{eqn:qr}
u_i=\boldsymbol{\beta}_\tau^T\boldsymbol{g}_i ,
\qquad i=1,\ldots,n_y,
\end{equation}
with coefficients $\boldsymbol{\beta}_\tau$ estimated by minimising
\begin{equation}\label{eqn:qs}
\mathcal{R}_\tau
=\frac{1}{n_y\,n}\sum_{i=1}^{n_y}\sum_{d=1}^{n}\rho_\tau\bigl(y_{i,d}-u_i\bigr),
\end{equation}
where $\rho_\tau$ is the check function
\begin{equation}\label{eqn:check}
	\rho_\tau(z)=
		\begin{cases}
		    \tau\,z & \text{if } z\ge 0,\\
			(\tau-1)\,z & \text{if } z<0 .
		\end{cases}
\end{equation}
 
The scale and shape parameters are again modelled by Eq.~(\ref{eqn:sigma_xi}) and the dependence
parameter by Eq.~(\ref{eqn:alpha}), now based on the exceedances over the time-varying threshold.
Since $\phi=1-\tau$ is constant, the parameter vectors are
$\boldsymbol{\theta}_1=(\boldsymbol{\beta}_\tau,\boldsymbol{\beta}_{\sigma},\boldsymbol{\beta}_{\xi})$
and $\boldsymbol{\theta}_2=\boldsymbol{\beta}_{\alpha}$, again with $K=5$.
 
\subsection{Fitting procedure}
The Markov process model is fitted according to \citet{Smith_markov}. We follow \citet{Beirlant} and first fit the marginal and dependence structure separately by introducing the covariates into the parameter estimation, as mentioned before. The optimal degree for the logistic regression (constant threshold) and the quantile regression (time-varying threshold) is then determined by the Bayesian information criterion (BIC). Then the GPD is fitted as non-stationary in order to find the optimal degree of scale and shape parameter using the BIC. With the optimal degree of the Legendre polynomials for each parameter determined by the BIC, we fit the model jointly using the previously estimated parameters separately as starting values for the parameters, as suggested by \citet{Beirlant}. Here, ``jointly'' refers to joint optimisation of the GPD and dependence parameters. In
practice, Eq.~(\ref{eqn:optimisation}) is thus solved over
$(\boldsymbol{\beta}_{\sigma},\boldsymbol{\beta}_{\xi},\boldsymbol{\beta}_{\alpha})$, while
$\boldsymbol{\beta}_\phi$ (or $\boldsymbol{\beta}_\tau$, respectively) is held fixed at its
separately estimated value and enters as a plug-in estimate. The reason for this is that the year-to-year probability of exceedance (first approach) or the year-to-year threshold value (second approach) is unlikely to change because of minor adjustments to the dependence function. Some advantages of the joint fit are mentioned by \citet{Beirlant}, which are a better inference of the marginal parameters and a more reliable estimation of the dependence parameters.
 
\section{Attribution statements for ERA5 using ensemble simulations}
\label{sec:attribution_workflow}
 
\begin{figure*}
    \centering
    \includegraphics[width=0.8\linewidth]{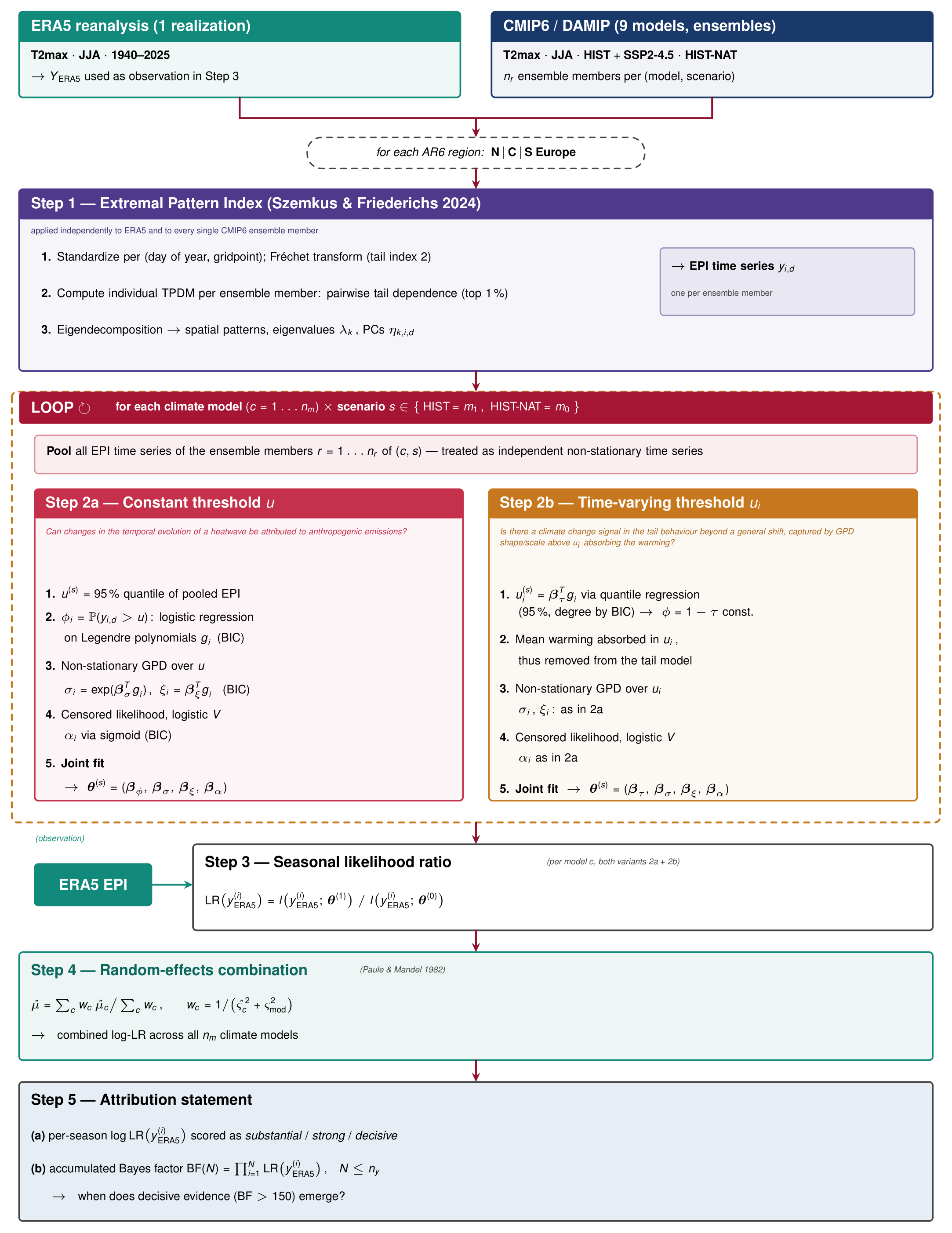}
    \caption{Flowchart of the attribution process, summarising Sect.~\ref{sec:attribution_workflow} and describing all necessary steps in order to obtain the attribution result.}
    \label{fig:flowchart_meurer2026}
\end{figure*}
 
The complete workflow of the attribution process is summarised in Fig.~\ref{fig:flowchart_meurer2026}.
 
\subsection{Model estimation with constant threshold}
\label{sec:model_estimation_with_constant_threshold}
To determine the likelihood, we need to estimate the parameter vectors for our scenarios $m_s$,
where $s=1$ refers to HIST and $s=0$ to HIST-NAT as described in Sect.~\ref{sec:data}. We
calculate the EPI for each realisation and both scenarios and pool the ensemble separately for
each climate model and scenario, assuming that the ensemble members $r=1,\ldots,n_r$ are
independent realisations of the non-stationary process under scenario $m_s$. We then estimate
the parameter vectors $\boldsymbol{\beta}_\phi^{(s)}$, $\boldsymbol{\beta}_{\sigma}^{(s)}$,
$\boldsymbol{\beta}_{\xi}^{(s)}$ and $\boldsymbol{\beta}_{\alpha}^{(s)}$, which yield the
exceedance probability $\phi^{(s)}_{i}$ over the threshold $u^{(s)}$, the GPD parameters
$\sigma^{(s)}_{i}$ and $\xi^{(s)}_{i}$, and the dependence parameter $\alpha^{(s)}_{i}$.
 
The parameter estimation process requires the implementation of a series of steps and the formulation of specific decisions.
\begin{enumerate}
    \item First, the threshold $u$ must be determined. It must be large enough to achieve asymptotic behaviour \citep[][Sect.~4.3.1]{Coles2001}, but small enough to allow sufficient data for reliable parameter estimation. In our application, we decided to use the 95~\% quantile of the respective ensemble simulations for the period 1940 to 2020 (HIST-NAT) and 1940 to 2025 (HIST \& SSP2-4.5). The threshold $u^{(s)}$ is therefore scenario-dependent. The advantage is that constant biases in the data are disregarded, and only the temporal evolution is investigated.
 
    \item In the next step, we estimate the time-varying exceedance probability $\phi^{(s)}_{i}$ above the constant threshold $u^{(s)}$ using logistic regression (Sect.~\ref{sec:using_constant_threshold}) for both scenarios.
    To determine the optimal maximum degree of the Legendre polynomials, we used the BIC. The fitting is performed using the statistical software package \verb|statsmodels| \citep{seabold2010statsmodels}. We model the non-stationarity of $\phi_{i}$ via its own regression formula because we expect this to be more robust than embedding it directly in the Markov likelihood.
 
   \item
   The non-stationary GPD is estimated with \verb|statsmodels| using maximum likelihood estimation. Again, the BIC is used to assess the optimal maximum degree of Legendre polynomials. Goodness-of-fit is assessed using quantile-quantile and probability-probability plots as described in \citet[Sect.~6.2.3]{Coles2001}. All combinations of scale and shape parameters up to degree $K=5$ are allowed, in which the degree of the shape parameter is not allowed to be larger than the degree of the scale parameter ($\text{deg}(\sigma)\geq \text{deg}(\xi)$). The reason for this condition is that the estimation of a covariate-dependent shape parameter introduces additional uncertainty \citep{friederichs2009}; to account for this, the shape parameter is here restricted so as not to vary more flexibly than the scale parameter. This improves the stability of likelihood-based inference. Here, we use all threshold exceedances and do not apply a declustering scheme.
 
    \item
    Next, the parametric form of $V$ must be specified, in this example, the logistic model. For each scenario, the univariate estimates $\phi^{(s)}_{i}$ and the optimal degrees of $\sigma^{(s)}_{i}$ and $\xi^{(s)}_{i}$ are plugged into the censored likelihood approach. Then the optimal degree of the covariates for the dependence parameter is determined. Finally, using the estimated parameters as starting values, the model is fitted jointly and the resulting parameters are returned by the fit. The logistic-regression coefficients are not updated in this joint fit. 
\end{enumerate}
 
\subsection{Likelihood of ERA5 with constant threshold}
The likelihood ratios of Sect.~\ref{sec:theory_attribution} are evaluated for the ERA5 EPI time
series. We write $\boldsymbol{Y}_{\mathrm{ERA5}}=(\boldsymbol{y}^{(1)}_{\mathrm{ERA5}},\ldots,
\boldsymbol{y}^{(n_y)}_{\mathrm{ERA5}})$ for the complete record and
$\boldsymbol{y}^{(i)}_{\mathrm{ERA5}}$ for the season of year $i$. The threshold
$u_{\mathrm{ERA5}}$ is the 95~\% quantile of the complete ERA5 EPI record over 1940 to 2025
(JJA), whereas the scenario parameters $\phi^{(s)}_{i}$,
$\sigma^{(s)}_{i}$, $\xi^{(s)}_{i}$ and $\alpha^{(s)}_{i}$ vary from season to season. The
scenario parameters describe the temporal evolution relative to the respective scenario
threshold $u^{(s)}$; applying them to the ERA5 threshold thus disregards constant biases
between the simulations and the reanalysis, consistent with the threshold choice in
Sect.~\ref{sec:model_estimation_with_constant_threshold}.

The evidence contributed by a single summer is then the seasonal likelihood ratio
\begin{equation}\label{eqn:probratio1}
\mathrm{LR}\bigl(\boldsymbol{y}^{(i)}_{\mathrm{ERA5}}\bigr)
=\frac{l\bigl(\boldsymbol{y}^{(i)}_{\mathrm{ERA5}};\,u_{\mathrm{ERA5}},\,
       \phi^{(1)}_{i},\sigma^{(1)}_{i},\xi^{(1)}_{i},\alpha^{(1)}_{i}\bigr)}
      {l\bigl(\boldsymbol{y}^{(i)}_{\mathrm{ERA5}};\,u_{\mathrm{ERA5}},\,
       \phi^{(0)}_{i},\sigma^{(0)}_{i},\xi^{(0)}_{i},\alpha^{(0)}_{i}\bigr)} ,
\end{equation}
where the seasonal likelihood $l(\boldsymbol{y}^{(i)};\cdot)$ is given by
Eq.~(\ref{eqn:markov_chain}). Because the seasons are assumed independent, the likelihood ratio
of the complete record follows as the product over seasons and coincides with the Bayes factor
of Eq.~(\ref{eqn:BayesII}),
\begin{equation}\label{eqn:probratio1_total}
\mathrm{LR}\bigl(\boldsymbol{Y}_{\mathrm{ERA5}}\bigr)
=\prod_{i=1}^{n_y}\mathrm{LR}\bigl(\boldsymbol{y}^{(i)}_{\mathrm{ERA5}}\bigr)=\mathrm{BF}.
\end{equation}
Equation~(\ref{eqn:probratio1}) therefore yields an attribution statement for a single summer,
and Eq.~(\ref{eqn:probratio1_total}) for any period of consecutive summers up to a season to be
chosen.

\subsection{Model estimation with variable threshold}
\label{sec:model_estimation_with_varying_threshold}

The attribution statement with a variable threshold $u_i$ is different, since here we only
consider the distribution above $u_i$. The time-varying threshold is assumed to contain a large
part of the anthropogenic climate change signal, which is thus removed. Following
Sect.~\ref{sec:model_estimation_with_constant_threshold}, the likelihood of $\boldsymbol{Y}$
under scenario $m_s$ is determined by the parameter vectors $\boldsymbol{\beta}_{\tau}^{(s)}$,
$\boldsymbol{\beta}_{\sigma}^{(s)}$, $\boldsymbol{\beta}_{\xi}^{(s)}$ and
$\boldsymbol{\beta}_{\alpha}^{(s)}$, which yield the time-varying threshold $u^{(s)}_{i}$, the
marginal parameters $\sigma^{(s)}_{i}$ and $\xi^{(s)}_{i}$, and the dependence parameter
$\alpha^{(s)}_{i}$.
Again, we pool the ensemble separately for each climate model and scenario and describe the necessary steps and decisions for the model estimation.
\begin{enumerate}
    \item The first step is to perform the quantile regression of the ensemble simulations for a sufficiently high quantile, in our case the 95~\% quantile, as selected in Sect.~\ref{sec:model_estimation_with_constant_threshold}, leading to the time-varying thresholds $u^{(s)}_{i}$. Again, we make use of the BIC to derive the optimal maximum degree of the Legendre polynomials. The fitting is performed once again using the statistical software package \verb|statsmodels|.
    \item In principle, the steps are similar to those described in Sect.~\ref{sec:model_estimation_with_constant_threshold}, but now use time-varying thresholds to account for the effect of the mean increase in the time series. For fitting the non-stationary GPD, once again the package \verb|statsmodels| can be used, and the model evaluation is the same as mentioned in Sect.~\ref{sec:model_estimation_with_constant_threshold}. As for the logistic regression, the quantile regression is estimated using its own regression formula.
    \item Again, the parametric form of $V$ must be specified, in this example, the logistic model. For each scenario, the thresholds $u^{(s)}_{i}$ and the optimal degrees of $\sigma^{(s)}_{i}$ and $\xi^{(s)}_{i}$ are plugged into the censored likelihood approach to determine the optimal degree of the covariates for the dependence parameter. Finally, the model is fitted jointly with the parameters estimated separately as starting values. The quantile-regression coefficients are kept fixed and are not re-estimated in the joint fit.
\end{enumerate}

\subsection{Likelihood of ERA5 with variable threshold}
For the model with a variable threshold, the ERA5 threshold $u_{i,\mathrm{ERA5}}$ is obtained
from the 95~\% quantile regression of the complete ERA5 EPI record over 1940 to 2025 (JJA) and
therefore varies from season to season. Since the exceedance probability is constant by
construction, it does not enter the ratio. The seasonal likelihood ratio reads
\begin{equation}\label{eqn:probratio2}
\mathrm{LR}\bigl(\boldsymbol{y}^{(i)}_{\mathrm{ERA5}}\bigr)
=\frac{l\bigl(\boldsymbol{y}^{(i)}_{\mathrm{ERA5}};\,u_{i,\mathrm{ERA5}},\,
       \sigma^{(1)}_{i},\xi^{(1)}_{i},\alpha^{(1)}_{i}\bigr)}
      {l\bigl(\boldsymbol{y}^{(i)}_{\mathrm{ERA5}};\,u_{i,\mathrm{ERA5}},\,
       \sigma^{(0)}_{i},\xi^{(0)}_{i},\alpha^{(0)}_{i}\bigr)} ,
\end{equation}
and the corresponding Bayes factor over any period of consecutive summers again follows from
Eq.~(\ref{eqn:probratio1_total}).

\subsection{Uncertainty assessment}
\label{sec:uncertainty}

According to \citet{PACIOREK201869}, uncertainties arise from different sources. Uncertainty with respect to internal climate variability is generally assessed using large climate model ensembles, whereas ensembles with different climate models as used in this study further assess model uncertainty.
Additional uncertainty arises from the observations of the climate system, which, however, is generally small compared to other sources of uncertainty.

To account for sampling uncertainty, two approaches are often used: the delta-method and bootstrapping \citep{JEON201624}. In this study, we use the bootstrap method. For the bootstrap method, we proceed as follows for each scenario (repeating this for both scenarios of each climate model).
\begin{enumerate}
    \item Split the data set into annual parts.
    \item For each year, sample $n_r$ indices with replacement from $\{1,\ldots,n_r\}$, where $n_r$ is the number of ensemble members of the respective climate model and scenario.
    \item For each year, the ensemble members corresponding to the selected indices are used.
    \item All years are combined.
\end{enumerate}
Using this bootstrap method, the attribution is then repeated $n_B$ times for each of the climate models; in this study, we use $n_B=10$.

\subsection{Combining different likelihood ratios}
\label{sec:combining_lines_of_evidences}
Due to the use of different climate models, we obtain a set of different likelihood ratios for an event or a time series.
One may ask which likelihood ratio should be favoured or how a combined likelihood ratio can be calculated, since when communicating attribution results, the interest often lies in a specific value rather than a range of possible values. We follow an approach presented by \citet{ascmo-10-159-2024} that aims to combine different pieces of evidence (i.e., logarithmic probability ratios in their study) using an approach also used in meta-analyses and known as the random-effects model. This model is based on a paper published by \citet{PauleMandel1982}.

The main idea is to decompose the total variability into a contribution from natural variability
$\varsigma_{\mathrm{nat}}$ and a model uncertainty $\varsigma_{\mathrm{mod}}$, resulting in
$\varsigma^2_{\mathrm{tot}}=\varsigma^2_{\mathrm{nat}}+\varsigma_{\mathrm{mod}}^2$. For each
climate model $c=1,\ldots,n_m$, we have a best estimate $\hat{\mu}_c$ of the logarithmic
likelihood ratio and an estimate $\hat{\varsigma}_c$ of its standard deviation from the
bootstrap approach discussed in Sect.~\ref{sec:uncertainty}. The focus now lies on estimating
the weights $w_c$ for each climate model $c$, which represent the belief or confidence in the
specific model, given by
\begin{equation}
    w_c(\varsigma_{\mathrm{mod}})
    =\frac{1}{\hat{\varsigma}_c^{\,2}+\varsigma_{\mathrm{mod}}^{2}}.
\end{equation}
The best estimate of the logarithmic likelihood ratio over the $n_m$ climate models is then
given by a weighted average
\begin{equation}
    \hat{\mu}(\varsigma_{\mathrm{mod}})
    =\frac{\sum_{c=1}^{n_m} w_c(\varsigma_{\mathrm{mod}})\,\hat{\mu}_c}
          {\sum_{c=1}^{n_m} w_c(\varsigma_{\mathrm{mod}})},
    \label{eqn:optimal_mu}
\end{equation}
and is only a function of the model uncertainty or the representation error term
$\varsigma_{\mathrm{mod}}$. For the combined uncertainty, we can write

\begin{equation}
    \varsigma^2_{\mathrm{tot}}
    = \frac{1}{\sum_{c=1}^{n_m} w_c} \sum_{c=1}^{n_m}
      w_c\bigl(\hat{\varsigma}_c^{\,2}+\varsigma_{\mathrm{mod}}^{2}\bigr).
\end{equation}

The estimator of $\varsigma_{\mathrm{mod}}^2$ can be calculated using the methodology based on
\citet{PauleMandel1982}. The main idea behind this approach is that the variable $Q$, which is
the sum of the ratios of the two different estimates of variability, is distributed according to
a chi-squared distribution with $n_m-1$ degrees of freedom,
\begin{equation}
    Q(\varsigma_{\mathrm{mod}}) := \sum_{c=1}^{n_m}
    \frac{\bigl(\hat{\mu}_c - \hat{\mu}(\varsigma_{\mathrm{mod}})\bigr)^2}
         {\hat{\varsigma}_c^{\,2} + \varsigma_{\mathrm{mod}}^{2}}
    \sim \chi^2_{n_m-1} .
\end{equation}
The expected value of this sum is given by $\mathbb{E}[Q] = n_m - 1$, from which we can
estimate $\varsigma_{\mathrm{mod}}$ using the formula
\begin{equation}
    \hat{\varsigma}_{\mathrm{mod}} =
    \begin{cases}
        0, & \text{for } Q(0) \leq n_m - 1, \\
         \mathrm{Root}(Q - n_m + 1) , & \text{for } Q(0) > n_m - 1.
    \end{cases}
\end{equation}
Thus, the estimation of $\varsigma_{\mathrm{mod}}$ amounts to a root-solving problem if
$Q(0) > n_m - 1$.

The random-effects model of Eq.~(\ref{eqn:optimal_mu}) treats the climate models as a priori
independent. Given the documented dependence among CMIP6 models, which share, e.g., components or parameterisations, this assumption is likely to be violated to some
degree; the resulting $\varsigma_{\mathrm{tot}}$ should therefore be regarded as a lower bound
on the combined uncertainty.

\section{Results}\label{sec:results}

\subsection{Model parameter estimation}\label{sec:tasmax_attribution_estimation}
\begin{figure*}
    \centering
    \includegraphics[width=0.9\linewidth]{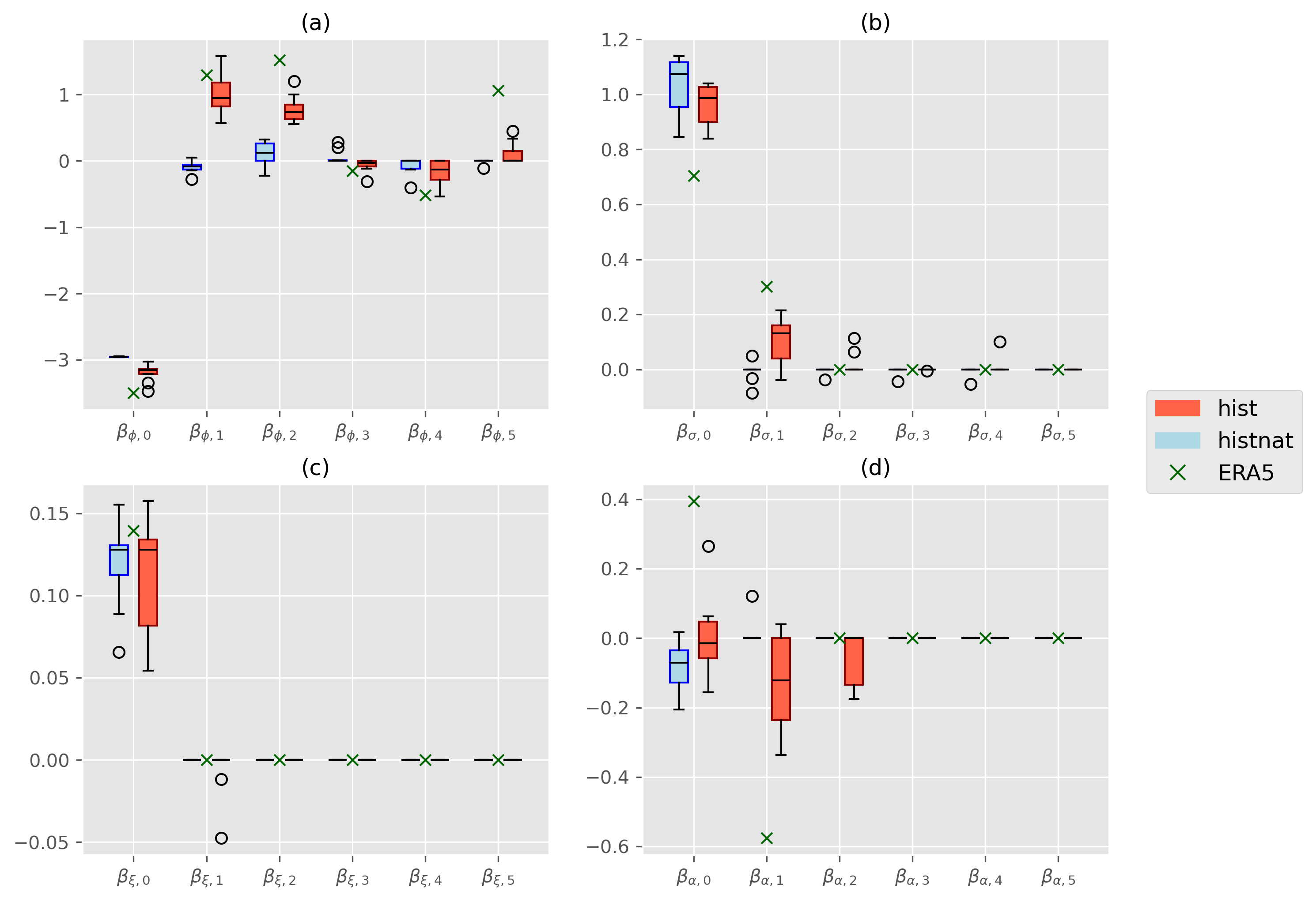}
    \caption{Estimated coefficients for the model with constant threshold (Sect.~\ref{sec:model_estimation_with_constant_threshold}) in the southern European region. The coefficients (x-axis) are shown for (a) the threshold exceedance $\phi_i$, (b) the scale $\sigma_i$, (c) the shape $\xi_i$, and (d) the dependence $\alpha_i$. The estimates for the different climate models are represented as box-whiskers, those for ERA5 as green crosses. The whiskers and fliers cover the whole range of data. }
    \label{fig:parameter_estimates_model1}
\end{figure*}

According to Sect.~\ref{sec:attribution_workflow}, both models with constant and time-varying thresholds are fitted to the data. For each climate model and scenario, we obtain a set of parameter coefficients.
Figure \ref{fig:parameter_estimates_model1} shows the coefficient estimates of the model with a constant threshold (Sect.~\ref{sec:model_estimation_with_constant_threshold}) for the EPI of the southern European region. The same model was also fitted to the EPI in ERA5 for comparison.
The first coefficient in each panel always represents the intercept, the second is a linear trend over time, and the third is a quadratic trend.
For the HIST-NAT scenario simulations, we do not expect tendencies, and indeed the intermodel variability of the trend parameters generally includes the zero line.

For the simulations of the HIST scenario, the exceedance probability shows significant positive trends, particularly in the linear and quadratic Legendre polynomials. For some CMIP6 models, the higher-order polynomials improve the BIC, but the signals are not consistent between the models. The parameter estimates for the threshold exceedance parameters in ERA5 are close to the range spanned by the estimates of the CMIP6 models. The uncertainty in the ERA5 parameter estimation is significantly larger than in the CMIP6 models, as the latter are estimated using a simulation ensemble and not, as in ERA5, a single realisation. Consequently, the ERA5 estimates may lie outside the uncertainty range given by the models.

The scale parameter estimates in the HIST scenario show a consistent linear increase, with a less consistent quadratic component. Again, ERA5 shows a similar behaviour in the non-stationarity of the scale parameter. However, the positive tendency in the scale parameter is counteracted by a negative linear trend in the shape parameter. This negative trend in the shape parameter is visible in some models in the HIST scenario, but not in ERA5. The overall positive shape parameter is expected in all scenarios, all simulations, and in ERA5, since the input data are the EPI, and the EPI in turn relies on Fr\'echet-transformed T2max values. For the dependence parameter, the linear and quadratic coefficients are relevant, while in ERA5 only the linear term is non-zero.
Similar tendencies in the parameters are observed for the northern and central European regions as shown in the Appendix (Sect.~\ref{sec:appendix_params}) in Figs.~\ref{fig:parameter_estimates_model1_tasmax_reg_16} and \ref{fig:parameter_estimates_model1_tasmax_reg_17}, with a similarly good agreement between the CMIP6 simulations and ERA5.

\begin{figure*}[t]
    \centering
    \includegraphics[width=0.9\linewidth]{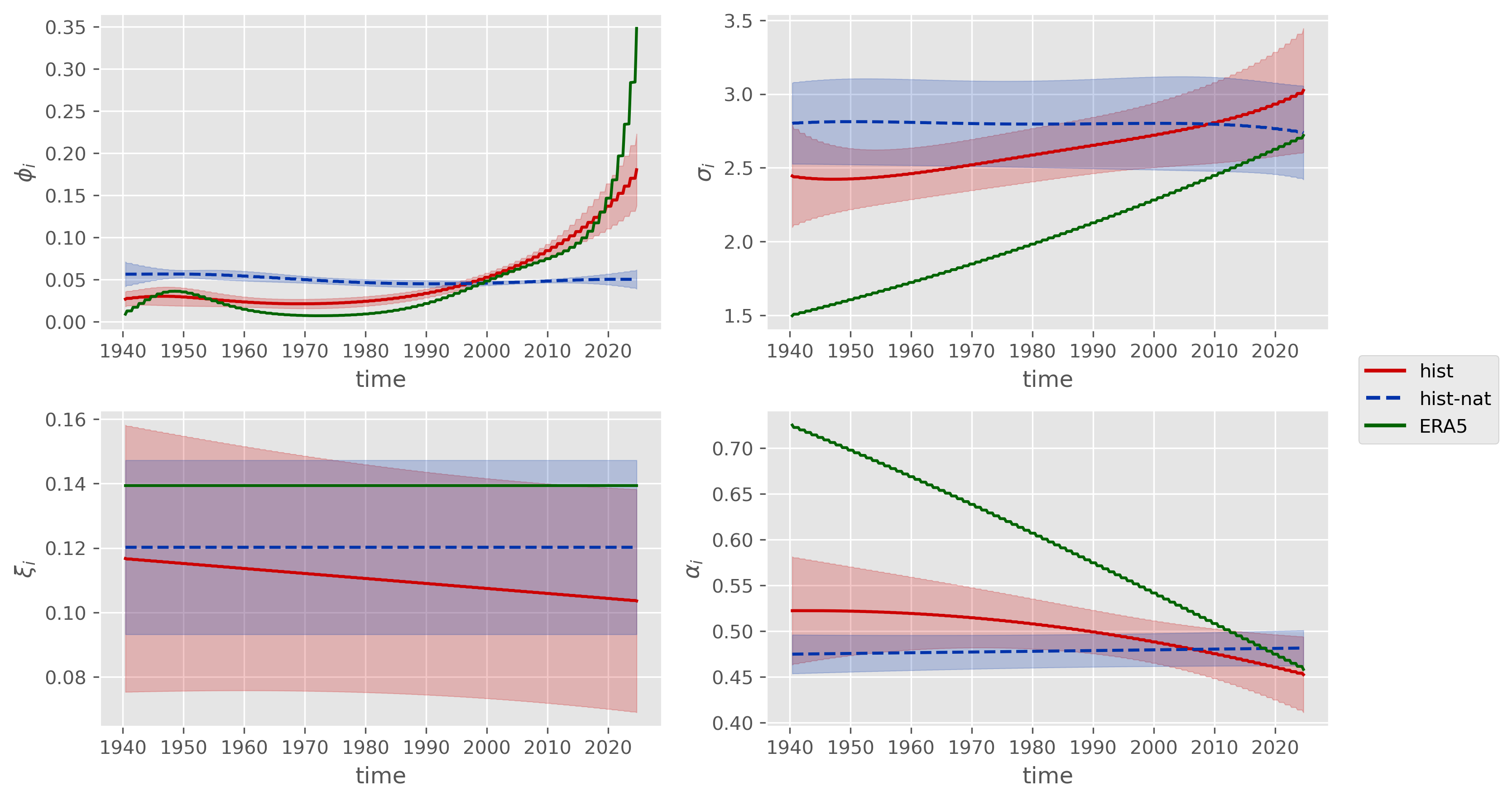}
    \caption{Temporal evolution of (a) threshold exceedance probability, (b) scale, (c) shape, and (d) dependence parameters for the model with constant threshold in the southern European region. Mean (solid line) and standard deviation (shading) for the different climate models are plotted based on the estimates in Fig.~\ref{fig:parameter_estimates_model1}. }
    \label{fig:all_parameters_reg_19_model1}
\end{figure*}

The resulting temporal evolution of the parameters is shown in Fig.~\ref{fig:all_parameters_reg_19_model1}. The increasing threshold exceedance probability in the HIST scenario is clearly visible, with a strong increase starting in the 1980s as a consequence of the second-order Legendre polynomial. A similar trend, with an even stronger increase during the last decade, is visible for the ERA5 threshold exceedance probability.
The scale parameter increases in the HIST scenario. Although this trend is stronger and starts at a lower level, it is consistent with the trend in ERA5. The trends in the shape parameter show large uncertainties between the different CMIP6 models. The ERA5 shape parameter is at the upper uncertainty level of the models and could counteract the effect of the comparatively low scale parameter.

The increasing dependence (i.e., decreasing $\alpha_i$) is visible for the HIST scenario and ERA5.
The dependence in ERA5 is weaker at the beginning, and the trend towards increasing dependence is significantly stronger compared to the HIST scenario. The increasing dependence reflects the greater frequency and duration of heatwaves, which leads to higher persistence of
hot extremes, and therefore, more consecutive hot days.

\begin{figure*}[t]
    \centering
    \includegraphics[width=0.9\linewidth]{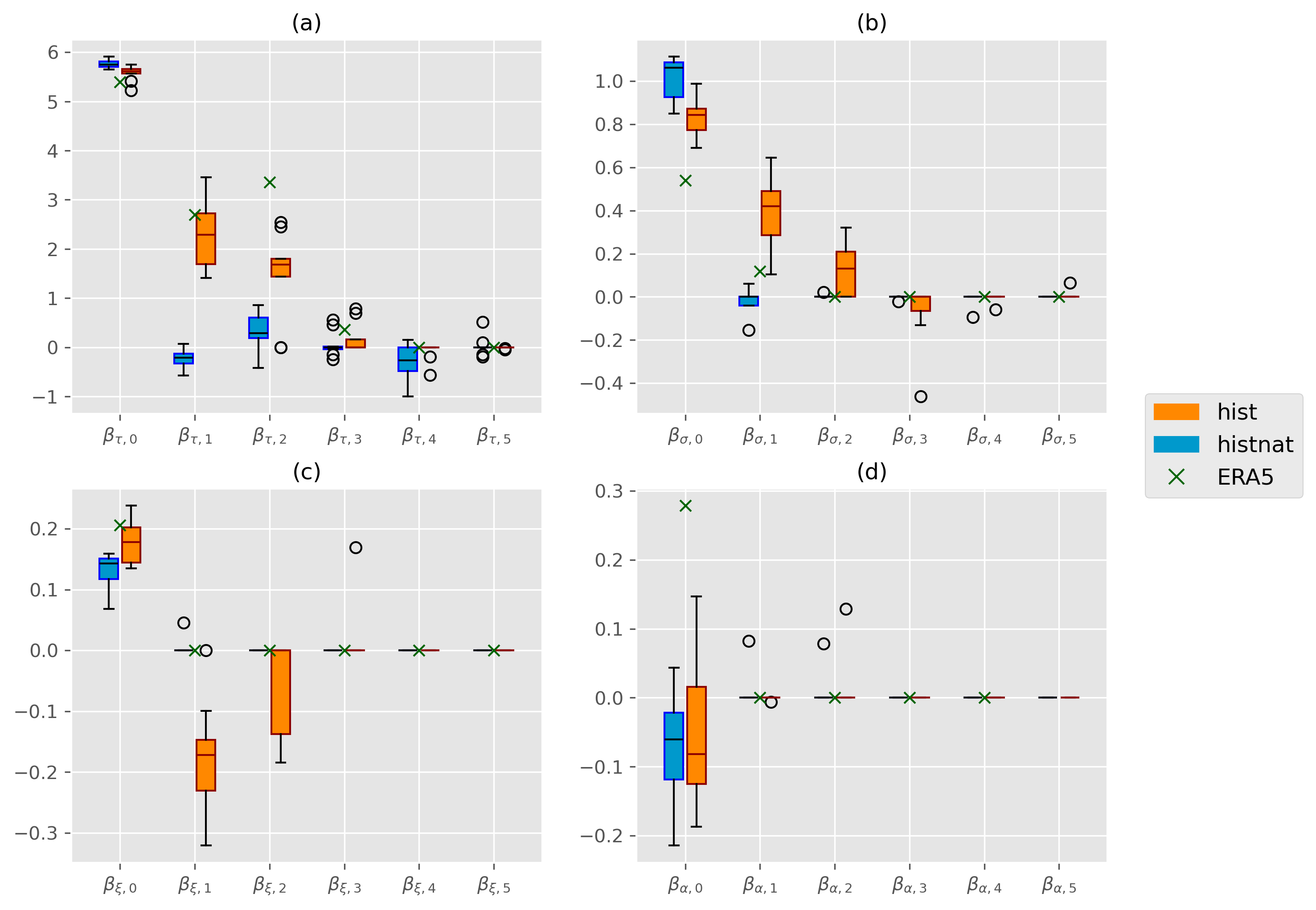}
    \caption{Same as Fig.~\ref{fig:parameter_estimates_model1} but for the model with time-varying threshold (Sect.~\ref{sec:model_estimation_with_varying_threshold}) in the southern European region, except that (a) shows the quantile regression parameters of the time-varying threshold.}
    \label{fig:parameter_estimates_model2}
\end{figure*}

Figure \ref{fig:parameter_estimates_model2} shows the coefficient estimates of the model with time-varying threshold (Sect.~\ref{sec:model_estimation_with_varying_threshold}) for the EPI of the southern European region. In the model with a time-varying threshold, there are no coefficients for the threshold exceedance probability, since this is constant by definition. Instead, we display the quantile regression parameters, which contain a large part of the non-stationarity but which are not part of the likelihood model.

For the HIST-NAT scenario, the model variability of the trend parameters includes the zero line, similar to the model with a constant threshold.
For the HIST scenario, the linear and quadratic coefficients of the quantile regression are significantly different from zero, and thus show a positive trend. ERA5 exhibits behaviour similar to that of the HIST scenario.

For the scale and shape parameters, the linear and quadratic coefficients become relevant. Although the coefficients indicate a positive trend in the scale parameter in the HIST scenario, the trend for the shape parameter is negative. Higher-order coefficients are only relevant for a subset of the climate models in the HIST scenario. In contrast, for ERA5 only the first-order Legendre polynomial becomes relevant for the scale parameter, while the shape parameter is stationary.
In terms of dependence, trend coefficients are only relevant for a small subset of models. This indicates that stationary dependence is the norm for both scenarios, as is the case for ERA5.

\begin{figure*}[t]
    \centering
    \includegraphics[width=0.9\linewidth]{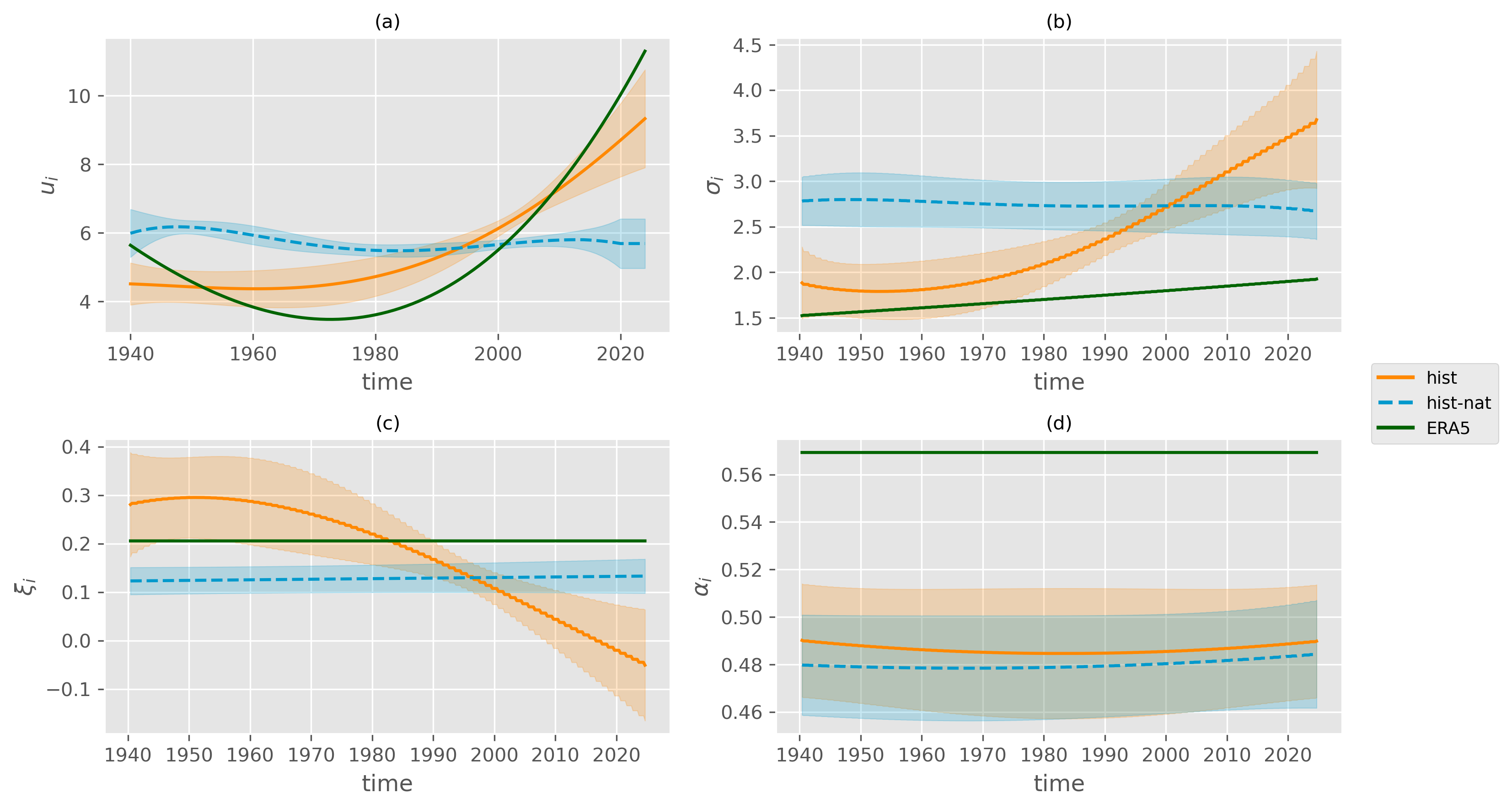}
    \caption{Same as Fig.~\ref{fig:all_parameters_reg_19_model1} but for the model with time-varying threshold in the southern European region. }
    \label{fig:all_parameters_reg_19_model2}
\end{figure*}

The resulting temporal evolution of the parameter estimates (Fig.~\ref{fig:all_parameters_reg_19_model2}) of the HIST-NAT scenario does not show significant trends.
An increase in the threshold is visible for the HIST scenario and in ERA5, which is a consequence of the general warming trend.
In this HIST scenario, the scale parameter also increases with time. In contrast, the shape parameter initially increases slightly, then decreases, and becomes negative for most of the climate models after 2010. This behaviour is not reflected by ERA5.
Regarding the dependence, the temporal evolution is nearly constant for both scenarios and ERA5.

The negative shape parameter ($\xi_i<0$) in the HIST scenario means that the exceedances are bounded above, with an upper limit of $u_i-\sigma_i/\xi_i$ \citep[Sect.~4.2.1]{Coles2001}. We therefore observe that the distribution of exceedances of the time-dependent 95~\% quantile becomes light-tailed over time.
This effect is partially offset by the increases in the scale parameter. It can thus be concluded that although the variance of the exceedances increases, the probability of large outliers decreases.
However, the tendency towards a negative shape parameter may also indicate that the Fr\'echet transformation required to determine the EPI is no longer adequate and that the strong non-stationarity must therefore be taken into account when defining the spatial patterns.

What we can show is that the sign change of the shape parameter in this model shows a significant improvement in BIC compared to a model that either keeps the shape parameter constant or restricts it to the positive value range.
Very similar tendencies, particularly the negative trend in the shape parameter, are also observed for the other regions, as shown and described in the Appendix in Sect.~\ref{sec:appendix_params} in Figs.~\ref{fig:parameter_estimates_model2_tasmax_reg_16} and \ref{fig:parameter_estimates_model2_tasmax_reg_17}.

Regarding potential biases of the GCMs, we can assume that first- and second-order biases are removed by the standardisation to Fr\'echet. This ensures a common scale for all the different climate models. Another bias is removed by standardising the EPI. A good indicator for this is that the ERA5 parameter estimates generally lie within the intermodel spread of the HIST scenario. Due to the smaller sample size, the sampling uncertainty of ERA5 and of the individual models is much larger, but it is not displayed in the figures. This might explain some of the discrepancies where the ERA5 coefficients fall outside the uncertainty range of the HIST coefficients. If the estimates from ERA5 and the climate models differ significantly, this may indicate differences in the physical processes represented in the models. However, this investigation is beyond the scope of this work. Better climate models with higher resolution may be required. It could also point to inconsistent changes in ERA5. The advantage of our modelling approach is that such differences can be made visible.

\subsection{Attribution results}\label{sec:tasmax_attribution}
\begin{figure*}[t]
    \centering
    \includegraphics[width=1\linewidth]{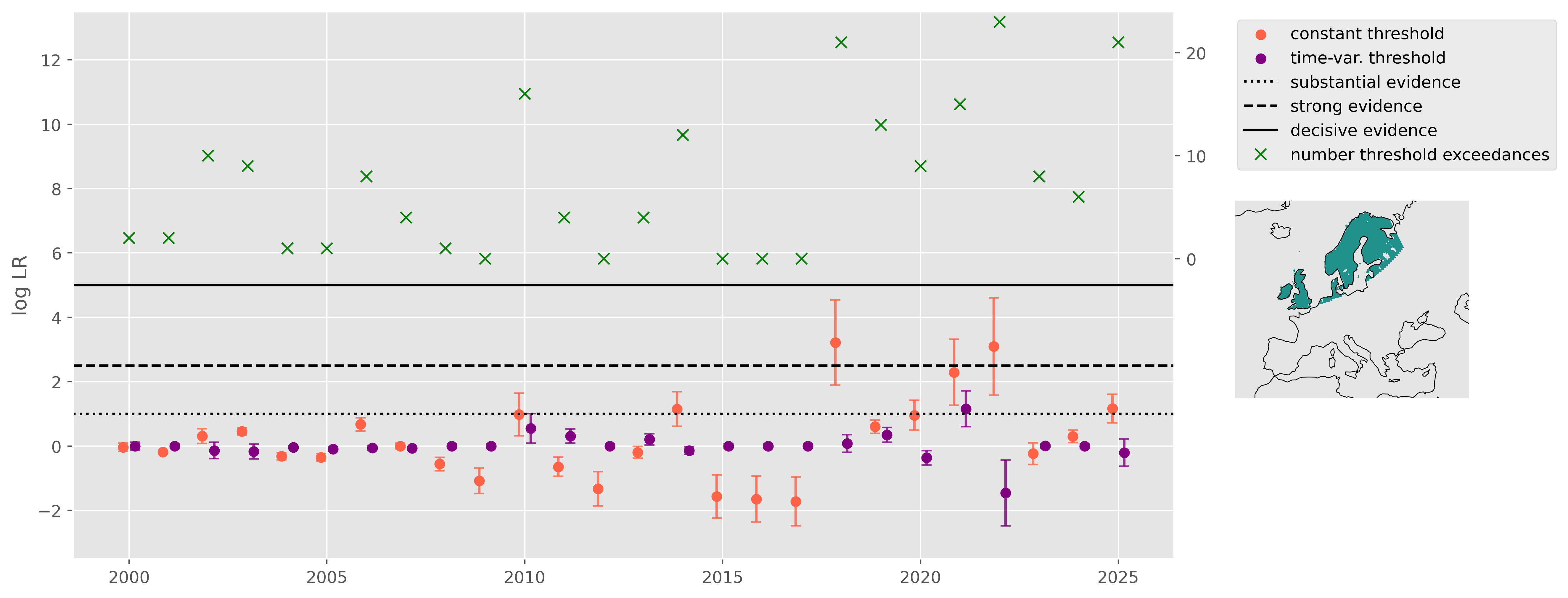}
    \caption{Logarithmic seasonal likelihood ratio $\log\mathrm{LR}(\boldsymbol{y}^{(i)}_{\mathrm{ERA5}})$ including $\pm$ standard deviation (y-axis left) for the northern European region, estimated according to Sect.~\ref{sec:combining_lines_of_evidences}. Green crosses indicate the number of threshold exceedances of the ERA5 EPI over the 95~\% quantile in the respective summer (y-axis right).}
    \label{fig:boxplot_probability_ratio_reg_16}
\end{figure*}

\begin{figure*}[t]
    \centering
    \includegraphics[width=1\linewidth]{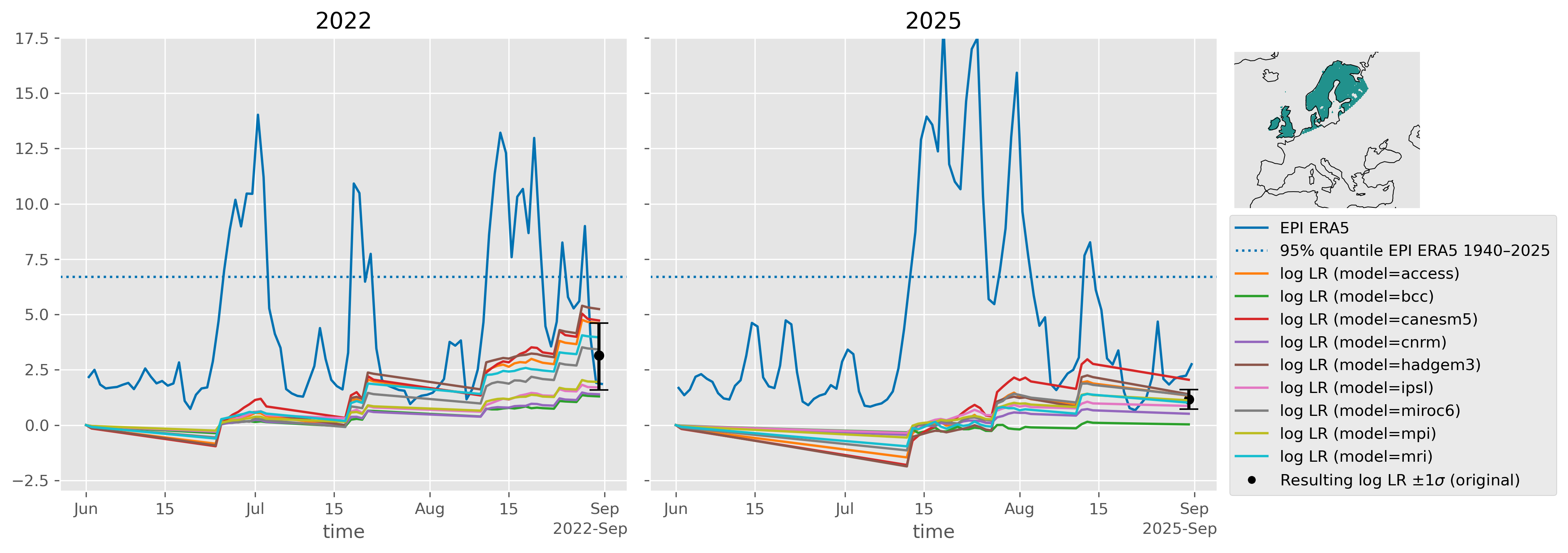}
    \caption{EPI of ERA5 and the logarithmic seasonal likelihood ratio for the summers 2022 and 2025 in the northern European region and the constant threshold model. The estimate shown in Fig.~\ref{fig:boxplot_probability_ratio_reg_16} is marked in black. }
    \label{fig:summer_compare}
\end{figure*}

Given the parameter estimates of the non-stationary Markov models, we can calculate the seasonal likelihood ratios in Eq.~(\ref{eqn:probratio1}) and (\ref{eqn:probratio2}) according to Sect.~\ref{sec:combining_lines_of_evidences}.
Figure \ref{fig:boxplot_probability_ratio_reg_16} shows the logarithmic likelihood ratios for both models with constant and variable thresholds, and the respective standard deviations for summers in the northern European region since 2000. The first event that provided substantial evidence for the HIST scenario was in 2010, and two events in 2018 and 2022 even showed strong evidence for a single northern European summer time series. All events are documented as extreme heat events. The likelihood ratio for the summer of 2025 is not among the three highest likelihood ratios. This might be surprising, since the July 2025 heatwave was exceptional, as documented by \citet{report2025}. Our attribution refers to the entire summer period with a relatively cold June, which reduces the corresponding likelihood ratio (Fig.~\ref{fig:summer_compare}), while the attribution in \citet{report2025} only considers the hot phase in July. Compared to summer 2022, which has a similar number of threshold exceedances, the resulting likelihood ratio of summer 2022 is higher due to the different types of heatwave. In 2022, there were three heat events, whereas in 2025 there was only one (see Fig.~\ref{fig:summer_compare}), which affects the attribution result, since the Markov transitions are different. In general, we can conclude that the attribution result is affected not only by the number of threshold exceedances (as shown in Fig.~\ref{fig:boxplot_probability_ratio_reg_16}), but also by the type (short- or long-term events) of the event, the number of events as well as the strength of the EPI.

For the model with a time-varying threshold, in the northern European region, most summers have a logarithmic likelihood ratio of approximately zero (i.e., a likelihood ratio of one). This means that no other effect can be detected in the northern European region apart from a general increase in the threshold value. The significant difference in the likelihood ratios between 2021 and 2022 is due to the much higher EPI values observed during the heat events in 2021 compared to 2022.

\begin{figure*}[t]
    \centering
    \includegraphics[width=1\linewidth]{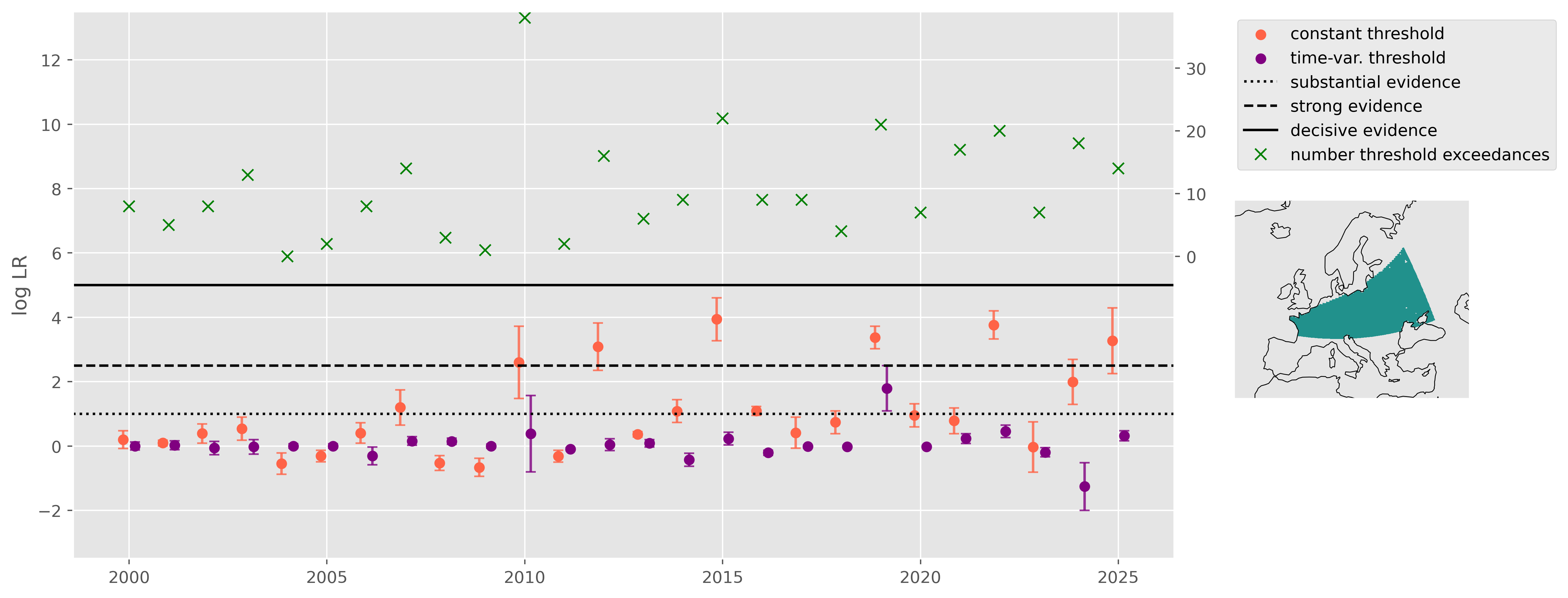}
    \caption{Same as Fig.~\ref{fig:boxplot_probability_ratio_reg_16} but for the central European region. }
    \label{fig:boxplot_probability_ratio_reg_17}
\end{figure*}

For the central European region, Fig.~\ref{fig:boxplot_probability_ratio_reg_17} shows more years with a logarithmic likelihood ratio that indicates strong evidence than for the northern European region. The summer of 2010 with the heatwave over eastern Europe and Russia is still outstanding in terms of threshold exceedances. The highest likelihood ratios are observed for the summers in 2010, 2012, 2015, 2019, 2022, and 2025. All of these years fall into the \textit{strong} category and are considered to be summers with heatwaves. The compound heat and drought event in 2018 \citep{nhess-25-541-2025} has a much smaller likelihood ratio, also with a lower number of threshold exceedances. In terms of maximum temperature, the summer of 2018 in central Europe was not as extreme as in the other years. The severe impacts in this summer were mainly due to the combination of high temperatures and drought.
The variable threshold model again provides no evidence beyond a trend in the threshold, except perhaps for the summer of 2019.

\begin{figure*}[t]
    \centering
    \includegraphics[width=1\linewidth]{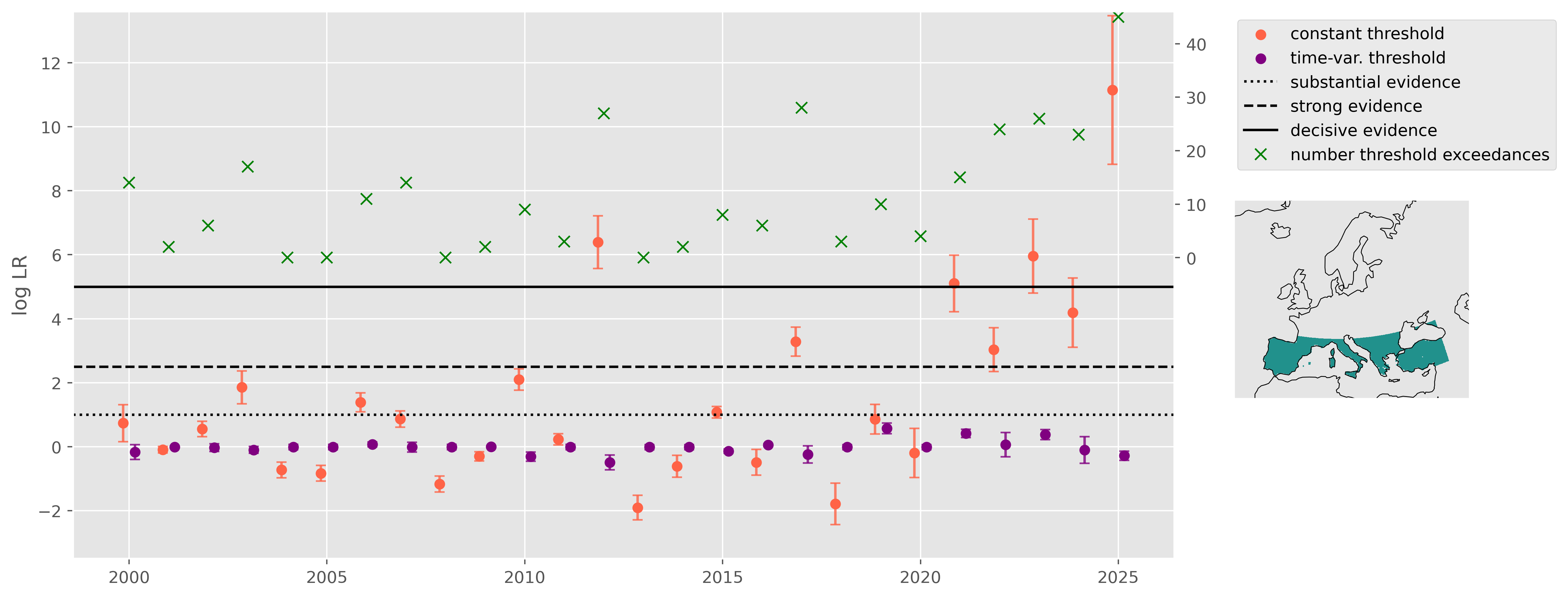}
    \caption{Same as Fig.~\ref{fig:boxplot_probability_ratio_reg_16} but for the southern European region.}
    \label{fig:boxplot_probability_ratio_reg_19}
\end{figure*}
In the southern European region (Fig.~\ref{fig:boxplot_probability_ratio_reg_19}), the year 2025 was outstanding compared to the other years. On almost half of the summer days, the EPI for the southern European region exceeded the 95~\% quantile, whereas the summer was similar to the summer of 2024 in terms of average maximum summer temperatures. High likelihood ratios can be observed primarily since 2021, with the summers of 2012, 2021, 2023, and 2025 falling into the \textit{decisive} category. The southern European region clearly shows the strongest evidence of climate change among the three regions.

When the effect of the variable threshold is included, the logarithmic likelihood ratio fluctuates around zero in all regions. However, uncertainty appears to increase with higher threshold values, which can cause both downward and upward excursions over a period of several years. The negative shape parameter of the variable threshold model observed in the anthropogenic-driven simulations after 2010 is not observed in ERA5 and could therefore be an artefact of the numerical models. Alternatively, this could indicate possible changes in processes that are not yet apparent in the ERA5 data but occur in climate simulations under the driving scenario. Therefore, it is also possible that the modelled trends in the CMIP6 simulations are not yet observable in ERA5.

Years with a high likelihood ratio in the time-dependent threshold model are usually years with many event/non-event transitions, where the EPI exceeds the threshold on only one of two consecutive days. Such events become more likely in the HIST scenario, especially due to the change in the shape parameter. This applies to all three regions, as the parameters and their trends are similar for the different regions.

Finally, we are interested in the question of when in the past we would have had sufficient evidence for climate change using the means available to us today. To answer this question, we accumulate the seasonal likelihood ratios of Eq.~(\ref{eqn:probratio1}) over consecutive summers. Under uniform prior probabilities, Eq.~(\ref{eqn:BayesII}) gives

\begin{equation}\label{Eq:LR}
     \frac{\mathbb{P}\bigl(m_1\mid \boldsymbol{y}^{(1)}_{\mathrm{ERA5}}, \ldots ,
                                   \boldsymbol{y}^{(N)}_{\mathrm{ERA5}}\bigr)}
          {\mathbb{P}\bigl(m_0\mid \boldsymbol{y}^{(1)}_{\mathrm{ERA5}}, \ldots ,
                                   \boldsymbol{y}^{(N)}_{\mathrm{ERA5}}\bigr)}
     = \prod_{i=1}^{N} \mathrm{LR}\bigl(\boldsymbol{y}^{(i)}_{\mathrm{ERA5}}\bigr)
     = \mathrm{BF}(N) ,
\end{equation}
where $N\le n_y$ denotes the most recent season included, and $\mathrm{BF}(n_y)=\mathrm{BF}$
as defined in Eq.~(\ref{eqn:BayesII}). Varying $N$ from the first to the last
season allows us to follow the accumulation of evidence from 1940 to 2025. If $\mathrm{BF}(N)$
exceeds the value of 150 (see Table~\ref{tab:bayesfactor}), we conclude that there is decisive
evidence against the HIST-NAT scenario after the summer of season $N$.

\begin{figure*}[t]
    \centering
    \includegraphics[width=0.75\linewidth]{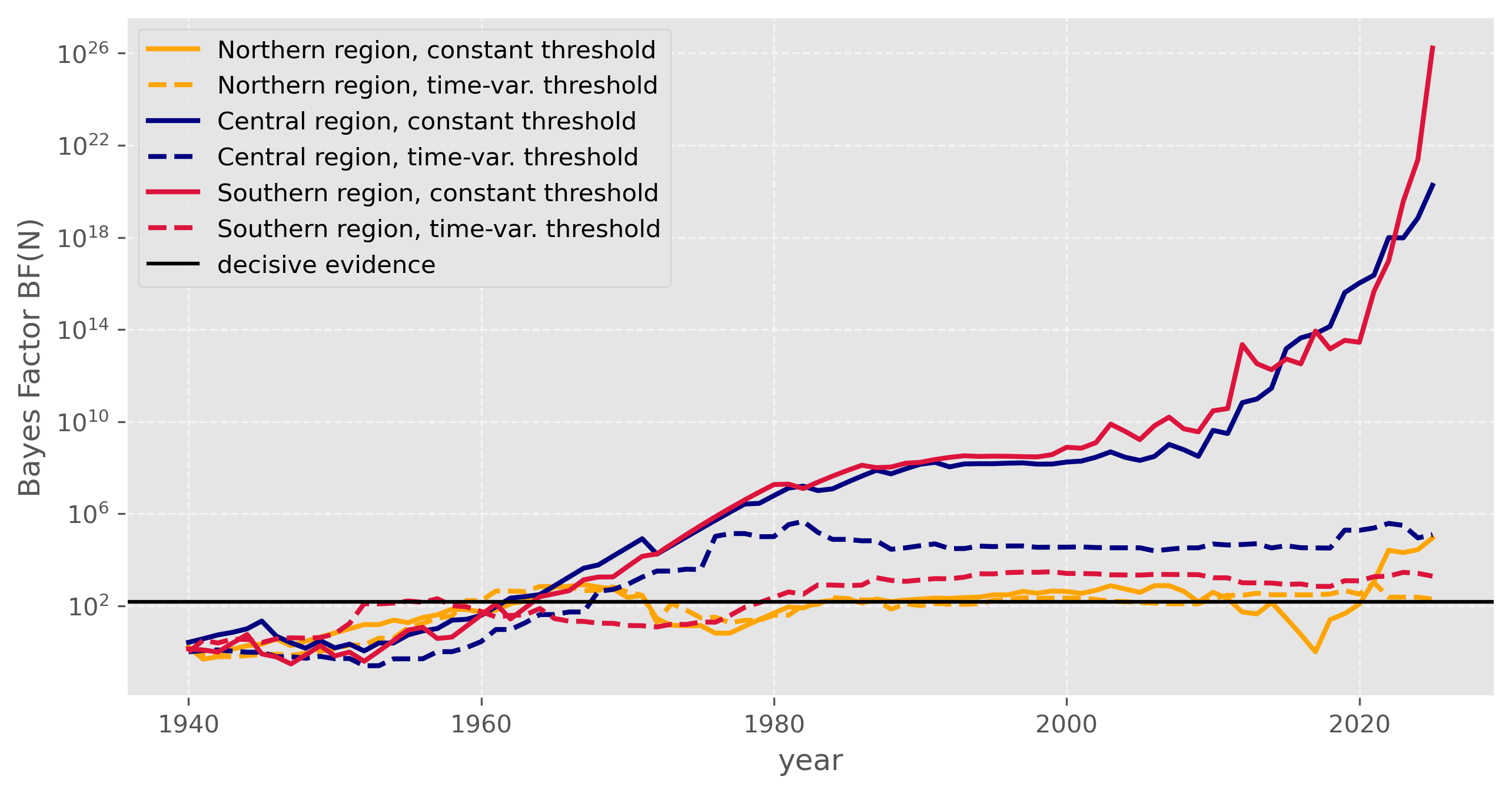}
    \caption{Accumulated Bayes factor $\mathrm{BF}(N)$ of Eq.~(\ref{Eq:LR}) for all summers since 1940 up to the most recent season $N$ included (x-axis, labelled by calendar year), assuming equal prior probabilities for both scenarios.}
    \label{fig:akkumuliert_pr_tasmax}
\end{figure*}

For the model with constant threshold, the level of decisive evidence is reached in 1962 in the central European region and in 1964 in the southern European region. For the northern European region, the Bayes factor has repeatedly fallen below and risen above this value since 1940. However, since 2021, it has remained consistently above 150.
For the central and southern European regions, the Bayes factor since 1940 has an increasing trend throughout the time period. For example, after the summer of 2025, we find a value of
\begin{equation}
    \mathrm{BF}(n_y)
    = \prod_{i=1}^{n_y}\mathrm{LR}\bigl(\boldsymbol{y}^{(i)}_{\mathrm{ERA5}}\bigr)
    \approx 2.25 \cdot 10^{26},
\end{equation} for the southern European region,
i.e. accumulated over the summers from 1940 to 2025.
In other words, the probability of the scenario with anthropogenic emissions given the summers from 1940 to 2025 is $10^{26}$ times higher than the probability of the scenario without anthropogenic emissions given the summers from 1940 to 2025.
For the model with a time-varying threshold, we can still attribute the entire period from 1940 onwards to the HIST scenario. As can be seen in Fig.~\ref{fig:akkumuliert_pr_tasmax}, after an overall increase in the Bayes factor from 1940 to 1980, it remains almost constant thereafter. This implies that summers after 1980 do not provide additional evidence for the HIST scenario in the model with a time-varying threshold, and all support for the HIST scenario is due to the period in which we do not expect an anthropogenic climate change signal.

\section{Discussion and conclusion}\label{sec:conclusion}

The Markov process model, which is based on bivariate extreme value theory, has been shown to be suitable for modelling the temporal dependence of the EPI, and thus deriving the likelihood of the observed EPI in ERA5 given different scenarios. Due to the Markov assumption, time periods ranging from two days to several years can be considered for attribution.
The non-stationarity of the Markov process allows us to examine the temporal changes in the time series due to climate change. To our knowledge, the adaptation of the censored threshold model to the non-stationary case is new. 
With a second model featuring a constant exceedance probability and a corresponding time-dependent threshold value, we can also answer the question of whether climate signals exist in the behaviour of extremes that go beyond a mean increase in the threshold value.

Our approach enables us to statistically model the temporal development of extreme events and thus also to conduct a corresponding attribution study. For heatwaves in Europe, there is a clear answer to the question of whether their temporal evolution can be attributed to anthropogenic emissions: The entire summer time series of the EPI from ERA5 can be assigned to the HIST scenario for each region with decisive evidence. Since 2000 in particular, heatwave events have increased in frequency, and the number of summers that can be attributed to the HIST scenario with strong or decisive evidence has risen. In the central and southern European regions, decisive evidence in favour of the scenario with anthropogenic emissions has been available since the 1960s, when considering all summers since 1940.

There is no clear answer to the second question, whether there is a climate change signal in the tail behaviour of heatwave extremes beyond a general shift in the temperature distribution. In the first years of the time series, an increasing shape parameter in the HIST scenario supports this hypothesis. However, the estimated non-stationarities over the last three decades significantly differ between the HIST scenario simulations and ERA5 in the variable threshold model. The trend towards a negative shape parameter in the HIST scenario suggests an upper endpoint for extremes, albeit counteracted by an increased scale parameter. In addition, the shape parameter is only slightly negative, so the upper bound applies only to very extreme observations. Neither is reproduced in ERA5.
Particular summers with many event/non-event transitions are attributed to the HIST scenario, since the respective CMIP6 model shape parameter estimates make these events more likely. All support for the HIST scenario is due to the period in which we do not expect an anthropogenic climate change signal.

What we can summarise is that heatwave extremes are becoming more extreme in terms of intensity — as evidenced by higher threshold exceedances over time.
However, once the general warming trend is absorbed into a time-varying threshold, the picture changes: in the HIST scenario, the GPD scale parameter increases, but there is no evidence of a corresponding increase in the shape parameter. In other words, relative to the shifting temperature distribution, the tail behaviour of exceedances has not changed in a detectable and consistent way across CMIP6 models and ERA5 -- and in this specific sense, extremes are not becoming more extreme. It is important to emphasise that this statement differs fundamentally from the first. The increasing frequency and intensity of heatwaves is a robust and decisive finding. In contrast, the absence of a detectable change in tail behaviour is a statement about the scale and shape of the distribution conditional on the warming trend having already been accounted for.

In principle, we can evaluate climate models using a similar approach as the World Weather Attribution. Unlike their analysis, we could take into account more than just linear trends \citep{Philip2020}, since the methodology for testing trends could also be applied to higher-order trend coefficients. Our results show that the models in the hist scenario essentially reproduce the ERA5 trends.
Despite the relatively complex statistical model, comparing the coefficients of the non-stationary model allows us to assess how well the past is represented in the climate models. However, this can only be done by taking into account the uncertainty in the ERA5 coefficient estimates. If non-stationarities in the historical datasets are well captured, this could also provide an indication of how well climate models can handle transient climate conditions.

In the Appendix, we further examine the influence of a seasonal cycle that may still be present in the data on the attribution (Appendix \ref{sec:seasonal_cycle}) and an extension to a second-order Markov process in Appendix \ref{sec:second_order}. Since both extensions have only a minor influence on the attribution statement, the results are not included in the main body of the article.

To summarise, the main contribution of this paper lies in the extension of the censored likelihood model for extremes to non-stationary time series, and its application to the attribution of the evolution of heatwaves over different regions in Europe. The significant non-stationarities in the temperature data are an issue for any extreme value model. A constant threshold model needs to be handled with care in the presence of strong temporal trends. In particular, the strong non-stationarity affects the estimated dependence structure in the constant threshold model.

Our second approach uses a time-varying threshold value to eliminate its effect and evaluate changes beyond it. However, the peaks-over-non-stationary threshold model of \citet{Friederichs2010} can also be adapted so that it can be used for extreme dependencies in time series and thus also, for example, for attribution in Eq.~(\ref{eqn:probratio2}). The main difference from \citet{Friederichs2010} in this article is that the time-varying threshold model uses a threshold estimated on ERA5 data to remove the ERA5 warming trends. The significantly stronger non-stationarities in future climate projections require sophisticated statistical extreme value models that adequately account for non-stationarities. This work is therefore an important step towards the statistical evaluation of non-stationary time series for extremes in the past and future. In this sense, our work can supplement existing analyses, such as those carried out by the World Weather Attribution, by offering a perspective that is less dependent on event definitions.

In terms of future work, we could also use models that combine asymptotic dependence and independence \citep[e.g.,][]{ramos_ledford} to attribute time series using the Markov process approximation.

\appendix
\label{sec:appendix}
\section{}    

\appendixfigures  

\subsection{Reasoning behind the non-stationary modelling of exceedance
probability and threshold}
\label{sec:reasoning}

In this subsection, we explain why we use non-stationary modelling for exceedance probability and threshold if the scale and shape parameters of the GPD are non-stationary. Throughout this subsection, we drop the day index $d$, in line
with assumption (\ref{eqn:no_seasonal}), and write $z\ge u$ for a generic level of interest;
the season index $i$ is dropped as well wherever the process is stationary.

Allowing the distributional parameters to depend on covariates already
induces non-stationary changes in the distribution -- but only with respect to the scale and shape
parameters of the GPD. Given a stationary process, we have, for
$z \ge u$,
\begin{equation}
    \mathbb{P}(y>z\mid y>u)
    =\left(1+\xi\,\dfrac{z-u}{\sigma}\right)^{-1/\xi}_{+}.
    \label{eqn:first}
\end{equation}
This is the conditional survivor function of observing a large value
$z$, given that the threshold
is exceeded \citep[Sect.~4.3.3]{Coles2001}. Only the scale and shape parameters of the GPD enter
here. An additional quantity that comes into play when considering the unconditional distribution
$\mathbb{P}(y>z)$ is the threshold exceedance probability
$\phi = \mathbb{P}(y>u)$. For the
censored threshold model, we require this unconditional distribution
(see Eq.~(\ref{eqn:marginal_gpd})), given by
\begin{equation}
    \mathbb{P}(y>z)
    =\mathbb{P}(y>u)\,\mathbb{P}(y>z\mid y>u)
    =\phi\left(1+\xi\,\dfrac{z-u}{\sigma}\right)^{-1/\xi}_{+},
    \qquad z \ge u. \label{eqn:uncond_stat}
\end{equation}
Equation~(\ref{eqn:uncond_stat}) is precisely the transformed margin
$v_{i,d}$ of Eq.~(\ref{eqn:margin}); the exceedance probability therefore enters the model
already at the stage of the transformation to standard Fr\'echet margins.

We now assume that the process is non-stationary and that a sequence of covariates
$\boldsymbol{g}_i$ is available; here these are Legendre polynomials. Equation
(\ref{eqn:first}) is then generalised to
\begin{equation}
    \mathbb{P}(y_i>z\mid y_i>u,\boldsymbol{g}_i)
    =\left(1+\xi_i\,\dfrac{z-u}{\sigma_i}\right)^{-1/\xi_i}_{+},
    \qquad z \ge u,
    \label{eqn:second}
\end{equation}
with $\sigma_i = \exp(\boldsymbol{\beta}_{\sigma}^T \boldsymbol{g}_i)$ and
$\xi_i = \boldsymbol{\beta}_{\xi}^T \boldsymbol{g}_i$,
as defined in Eq.~(\ref{eqn:sigma_xi}) of Sect.~\ref{sec:nonstat}, following
the covariate-dependent GPD framework of \citet{Eastoe}. As in
the stationary case, however, the censored likelihood requires the \emph{unconditional} distribution
of an exceedance, i.e.
\begin{equation}
    \mathbb{P}(y_i>z\mid \boldsymbol{g}_i)
    =\mathbb{P}(y_i>u \mid \boldsymbol{g}_i)\,
    \mathbb{P}(y_i>z\mid y_i>u,\boldsymbol{g}_i),
    \qquad z \ge u. \label{eqn:noncond}
\end{equation}
Because of the non-stationarity, the conditioning on $\boldsymbol{g}_i$ applies to \emph{all}
components of the model, including the exceedance probability, even though this conditioning is
often suppressed in the notation. We can therefore not use the stationary quantity
$\mathbb{P}(y>u)$, but must instead use the non-stationary threshold exceedance probability
$\phi_i=\mathbb{P}(y_i>u \mid \boldsymbol{g}_i)$,
which we model as in Eq.~(\ref{eqn:logit}),
\begin{equation}
    \mathrm{logit}(\phi_i)
    =\boldsymbol{\beta}_{\phi}^T \boldsymbol{g}_i,
    \label{eqn:logit2}
\end{equation}
using a logistic regression, again in line with \citet{Eastoe}.

The stationary quantity $\phi=\mathbb{P}(y>u)$ cannot simply be substituted for
$\mathbb{P}(y_i>u\mid\boldsymbol{g}_i)$ in Eq.~(\ref{eqn:noncond}).
The reason for this is that
Eq.~(\ref{eqn:noncond}) is not a modelling choice but an identity. Inserting a constant
$\phi$ in place of
$\phi_i=\mathbb{P}(y_i>u\mid\boldsymbol{g}_i)$
would violate this identity unless
$\mathbb{P}(y_i>u\mid\boldsymbol{g}_i)$ happened to be constant in time -- an assumption that is not
justified here, in particular given the underlying warming trend.

In other words, the two components are not two alternative ways of introducing non-stationarity;
they are two parts of the same non-stationary marginal model and cannot be specified independently
of one another. When the threshold $u$ is held fixed and the parameters
$\sigma_i$ and $\xi_i$ vary with $\boldsymbol{g}_i$,
the probability mass above $u$ varies as well.
Assuming a constant $\phi$ alongside time-varying $\sigma_i$ and $\xi_i$ would render the marginal model internally inconsistent.
Thus, the logistic regression in Eq.~(\ref{eqn:logit2}) is not an additional mechanism for
generating non-stationarity, but rather the component that maintains the coherence of the
fixed-threshold formulation. A natural alternative is to allow the threshold to vary, for example
as a time-varying quantile $u_i$ obtained by quantile regression
(Sect.~\ref{sec:using_time_varying_threshold}). This would hold the exceedance
probability constant by construction, shifting the non-stationarity into the threshold rather than
removing it.

\subsection{Including seasonal cycle in parameter estimation}
\label{sec:seasonal_cycle}
Although the seasonal cycle is removed when calculating the TPDM, a residual seasonal cycle remains in the EPIs, as can be demonstrated. One way to account for this is to model this residual seasonal cycle when fitting the parameters.

To this end, we relax the assumption of Eq.~(\ref{eqn:no_seasonal}) and allow the parameters to
vary within a season, so that the day index $d$ reappears. We assume that the residual seasonal
cycle can be described by sine and cosine functions and write
\begin{equation}\label{eqn:seasonal_covariates}
\boldsymbol{h}_d=\bigl(\sin(2\pi\,\mathrm{doy}_d/T_{\mathrm{yr}}),\ \cos(2\pi\,\mathrm{doy}_d/T_{\mathrm{yr}})\bigr)^T,
\qquad T_{\mathrm{yr}}=365.25,
\end{equation}
where $\mathrm{doy}_d$ is the day of the year corresponding to day $d$ of the season. While the
Legendre polynomials $\boldsymbol{g}_i$ of Sect.~\ref{sec:nonstat} describe the variation
between seasons, $\boldsymbol{h}_d$ describes the variation within a season. For the model with
a constant threshold (Sect.~\ref{sec:model_estimation_with_constant_threshold}), the parameters
then read
\begin{align}
\mathrm{logit}(\phi_{i,d})
 &= \boldsymbol{\beta}_\phi^T\boldsymbol{g}_i+\boldsymbol{\delta}_\phi^T\boldsymbol{h}_d,\\
\sigma_{i,d}
 &= \exp\bigl(\boldsymbol{\beta}_{\sigma}^T\boldsymbol{g}_i
              +\boldsymbol{\delta}_{\sigma}^T\boldsymbol{h}_d\bigr),\\
\xi_{i,d}
 &= \boldsymbol{\beta}_{\xi}^T\boldsymbol{g}_i+\boldsymbol{\delta}_{\xi}^T\boldsymbol{h}_d,\\
\alpha_{i,d}
 &= \bigl[1+\exp\bigl(-\boldsymbol{\beta}_{\alpha}^T\boldsymbol{g}_i
              -\boldsymbol{\delta}_{\alpha}^T\boldsymbol{h}_d\bigr)\bigr]^{-1},
\end{align}
with $\boldsymbol{\delta}_\phi$, $\boldsymbol{\delta}_\sigma$, $\boldsymbol{\delta}_\xi$ and
$\boldsymbol{\delta}_\alpha\in\mathbb{R}^2$ the additional coefficients to be estimated. For the
model with a time-varying threshold, the procedure is analogous. For the sake of simplicity, we do
not introduce a seasonal cycle into the quantile regression, i.e. $u_{i,d}\equiv u_i$ is
retained.

\begin{figure}[h]
    \centering
    \includegraphics[width=0.75\linewidth]{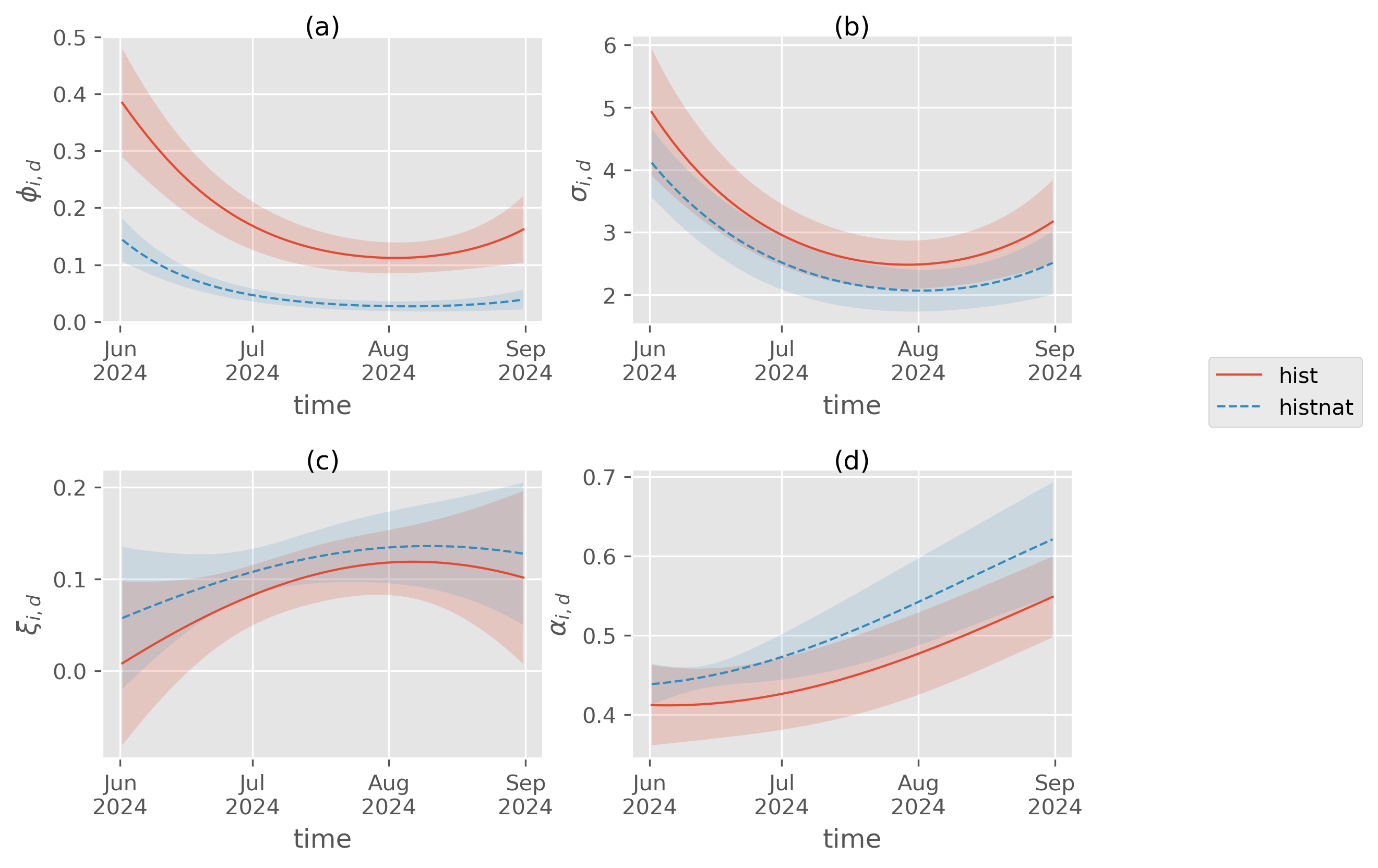}
    \caption{The seasonal cycle of (a) threshold exceedance probability, (b) scale, (c) shape, and (d) dependence parameters for the southern European region is shown for the year 2024, with mean and standard deviation calculated based on the different climate models.}
    \label{fig:seasonal_cycle_reg_19}
\end{figure}

\begin{figure}[h]
    \centering
    \includegraphics[width=0.75\linewidth]{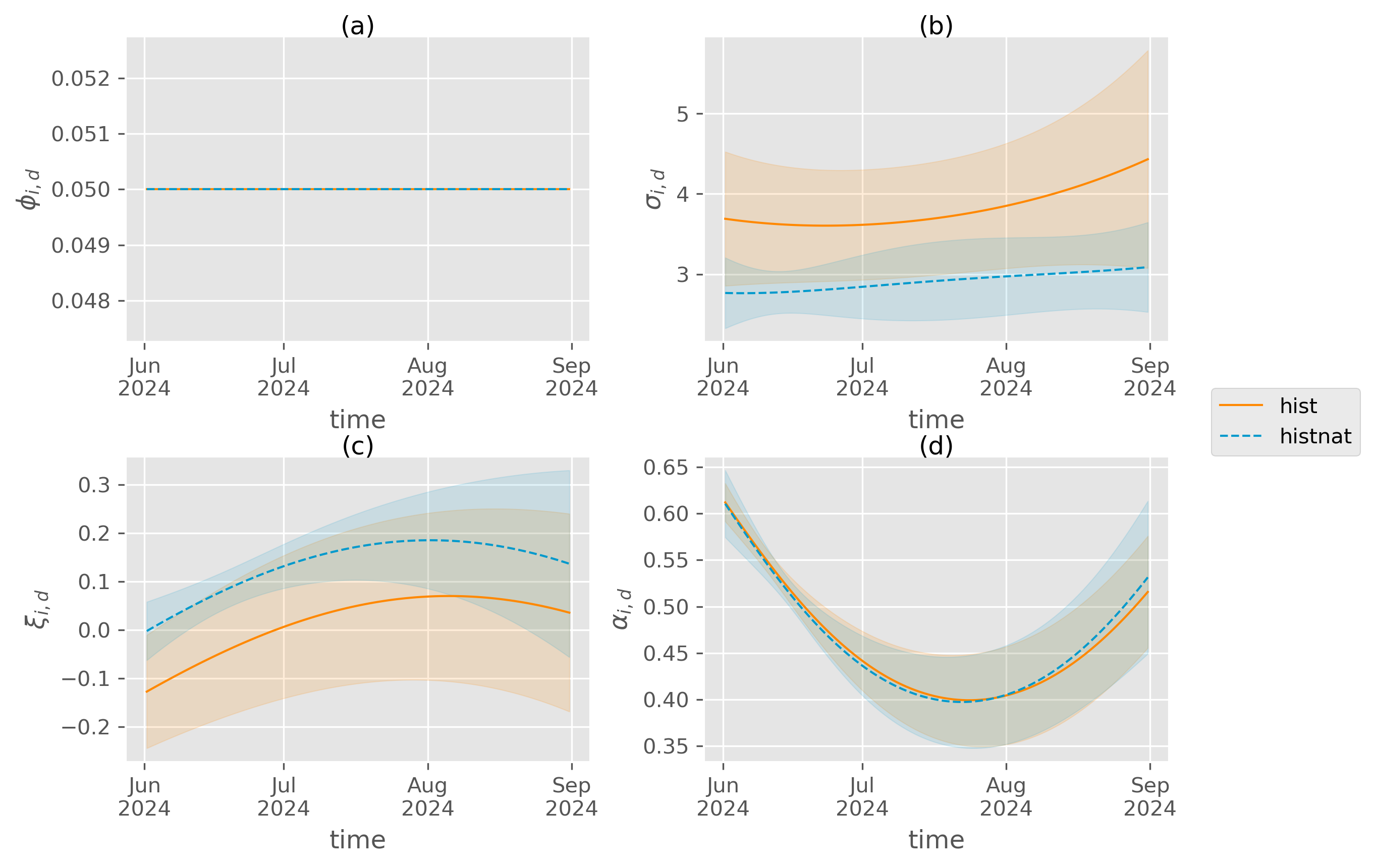}
    \caption{Same as Fig.~\ref{fig:seasonal_cycle_reg_19} but for the model with time-varying threshold.}
    \label{fig:seasonal_cycle_reg_19_model2}
\end{figure}

The workflow for fitting the models is then the same as described in Sect.~\ref{sec:attribution_workflow}.
Based on the maximum temperature example in the southern European region, the modelled seasonal cycle is plotted for models with constant and time-varying thresholds in Figs.~\ref{fig:seasonal_cycle_reg_19} and \ref{fig:seasonal_cycle_reg_19_model2}.

The seasonal cycle is similar for both models and scenarios. This suggests that the seasonal cycle may not shift significantly in response to climate change. However, the seasonal cycle shown here should be interpreted with caution, as it is a residual cycle remaining after the seasonal cycle has been removed from the data to calculate the TPDM.
To evaluate the effect of modelling the seasonal cycle on the estimation of the likelihood ratio, we calculate the seasonal likelihood ratio for each summer both with and without the seasonal cycle (as before).

\begin{figure}[h]
    \centering
    \includegraphics[width=0.65\linewidth]{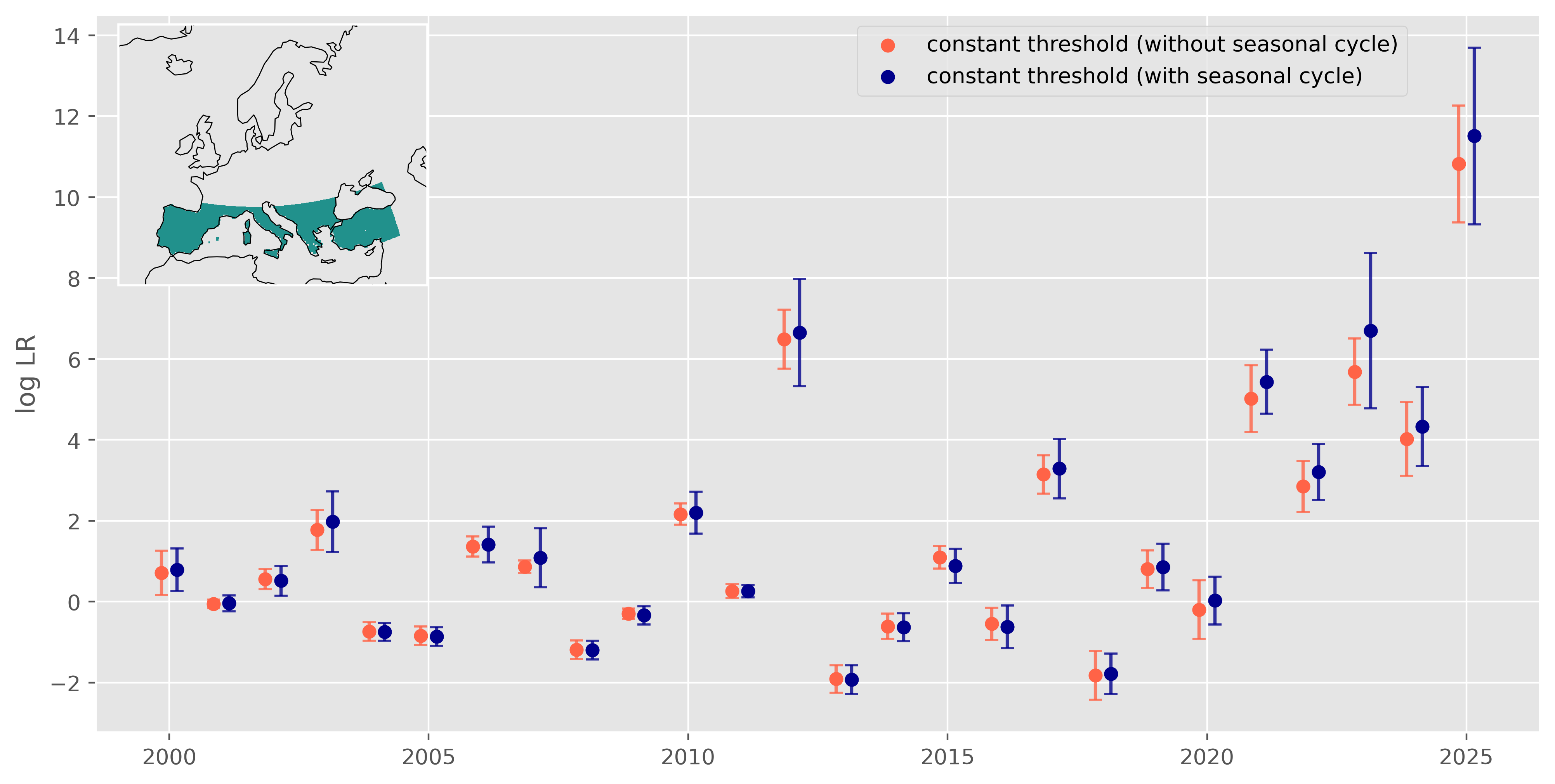}
    \caption{Logarithmic seasonal likelihood ratio including $\pm$ standard deviation for the southern European region, estimated according to Sect.~\ref{sec:combining_lines_of_evidences}, for the model with a constant threshold with and without modelling the seasonal cycle. For simplicity, no random effect model is used; instead, only the mean and standard deviation for the various models are displayed. }
    \label{fig:boxplot_probability_ratio_reg_19_seasonal_cycle_model1}
\end{figure}

The resulting likelihood ratios (Fig.~\ref{fig:boxplot_probability_ratio_reg_19_seasonal_cycle_model1}) are similar for many years, with slightly higher likelihood ratios (and greater uncertainty) observed for the last five years when modelling the seasonal cycle.
However, the effect of modelling the seasonal cycle is limited.

\subsection{Second-order Markov process}
\label{sec:second_order}

The effect of choosing a second-order Markov process will be shown based on maximum temperature and the southern European region, using a fitting procedure analogous to that of Sect.~\ref{sec:attribution_workflow} and the logistic model as tail dependence function, i.e. Eq.~(\ref{eqn:logistic_tail_3}) together with the likelihood of Eq.~(\ref{eqn:markov_chain_2}).

\begin{figure}[h]
    \centering
    \includegraphics[width=1\linewidth]{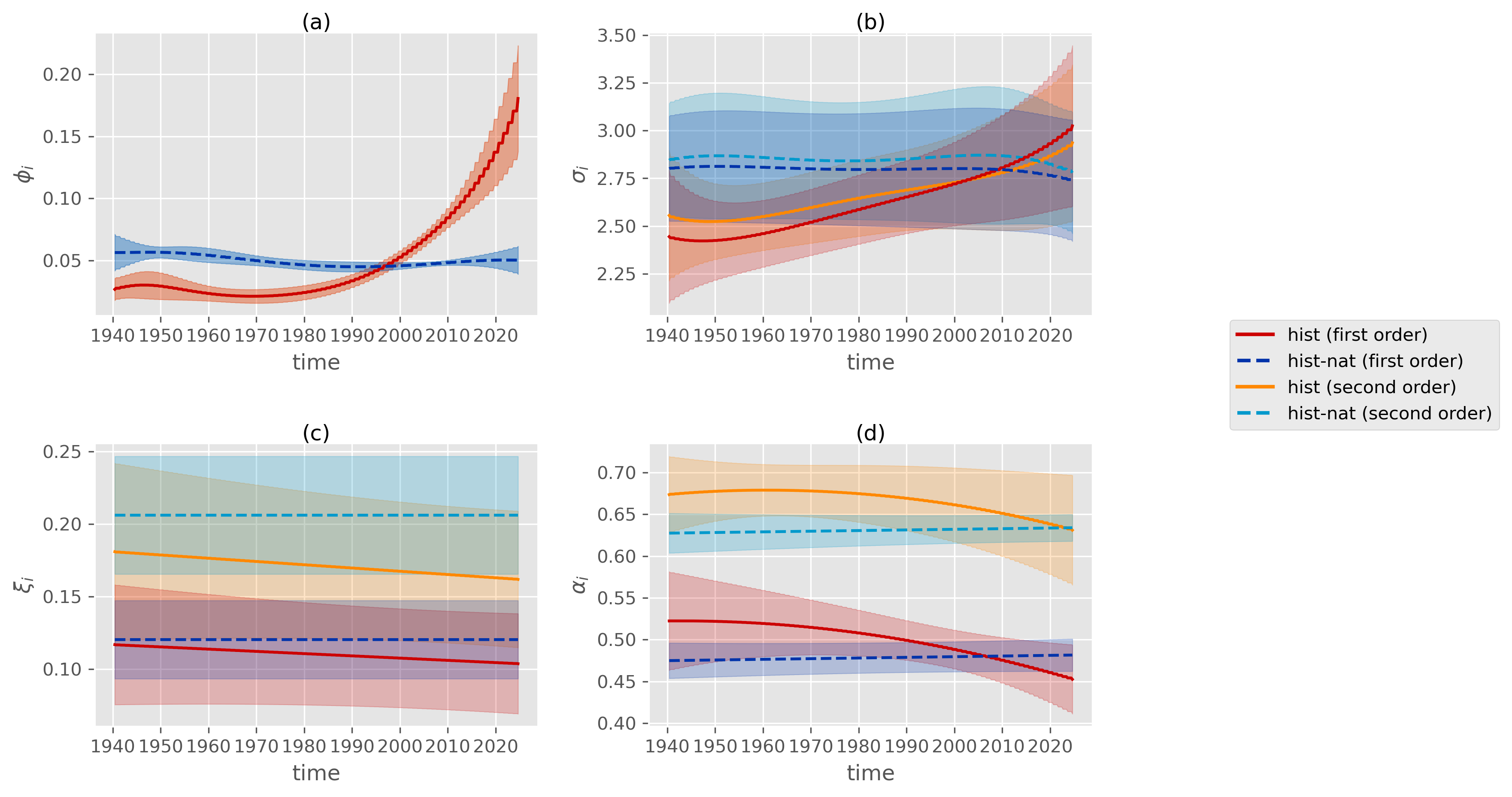}
    \caption{Same as Fig.~\ref{fig:all_parameters_reg_19_model1} but now including the parameter estimates for the second-order Markov process. }
    \label{fig:parameter_estimates_second_order}
\end{figure}

As can be seen in Fig.~\ref{fig:parameter_estimates_second_order}, the resulting scale parameters are in a similar range for the first- and second-order Markov process. The shape parameter estimates are slightly higher for the second-order Markov process in both scenarios, indicating a tendency towards greater variability in the data. As expected, the main difference between using a first- and second-order Markov process lies in the dependence parameter. Using the second-order Markov process results in weaker dependence. The reason is that, for higher time lags, the variability of the atmosphere reduces the dependence.

\begin{figure}
    \centering
    \includegraphics[width=0.75\linewidth]{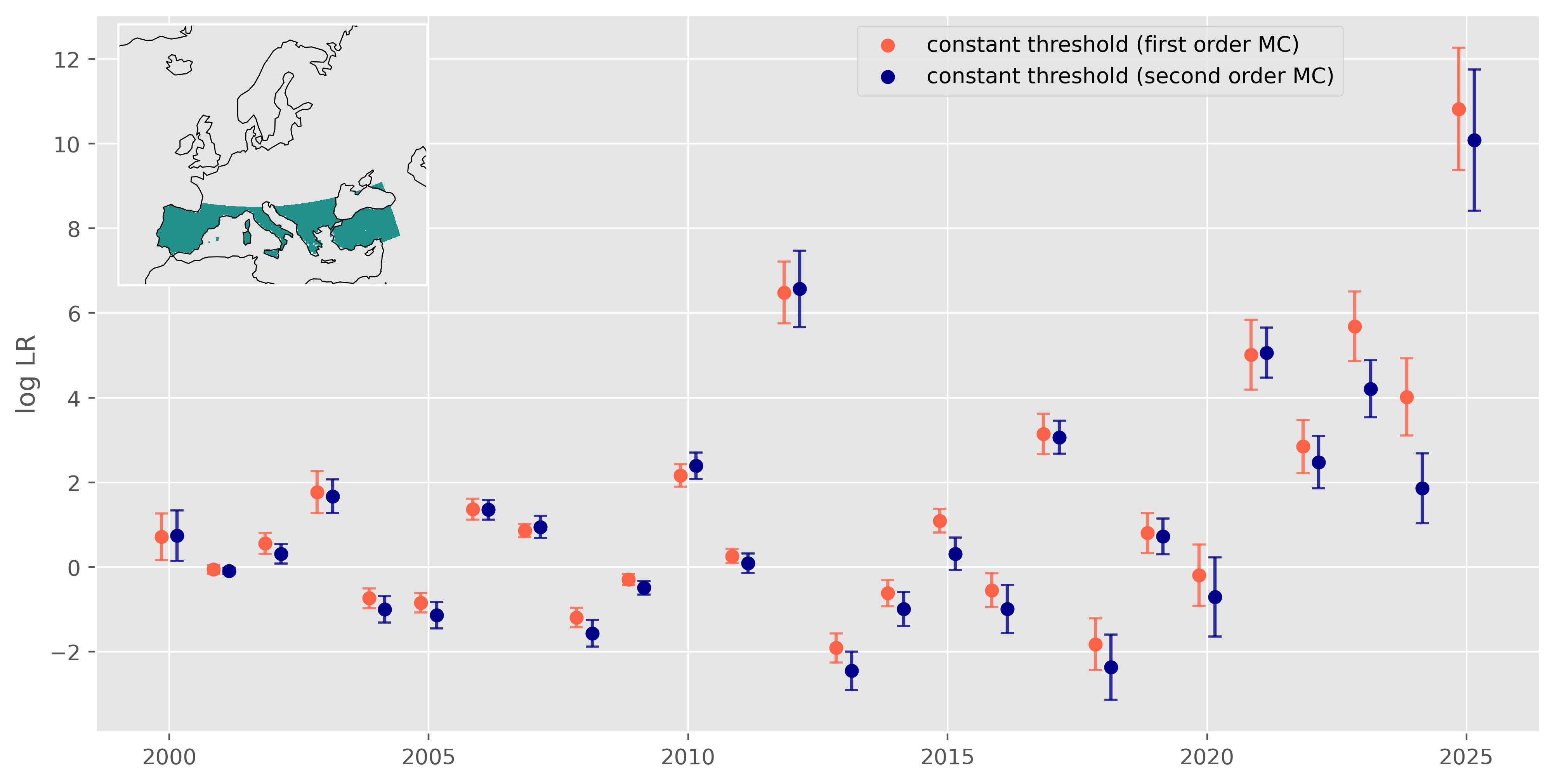}
    \caption{Effect of using a second-order Markov process: the mean and $\pm$ standard deviation of the seasonal likelihood ratios for the different climate models using a first- and a second-order Markov process. For the sake of simplicity, we did not apply the random effect model to combine the likelihood ratios of the different models. }
    \label{fig:boxplot_probability_ratio_reg_19_so_mc_model1}
\end{figure}

When calculating the likelihood ratios for the second-order Markov process model (Fig.~\ref{fig:boxplot_probability_ratio_reg_19_so_mc_model1}), a tendency towards smaller values can be seen for the second-order Markov process model over a large number of years. This may indicate the presence of dependencies with a lag greater than one day, such that the resulting likelihoods differ due to the dependence assumption of the first-order Markov process.

In order to assess whether the second-order Markov process gives a statistical improvement, the BIC (which is only one criterion pointed out by \citet{Smith_markov} for the model comparison) is compared. For both scenarios and all climate models, the second-order Markov process model gives an improvement in terms of the BIC. However, it is not straightforward and not necessarily possible to compare the Markov processes of different orders, due to the use of the censored likelihood \citep[Sect.~10.4.6]{Beirlant}.
In terms of attribution, we can conclude that choosing between a first- and second-order Markov process has only a small effect on the resulting likelihood ratio, and strong attribution to the HIST scenario is also present when using a second-order Markov process.

\subsection{Parameter estimates for the northern and central European regions}
\label{sec:appendix_params}
\begin{figure}[h]
    \centering
    \includegraphics[width=0.65\linewidth]{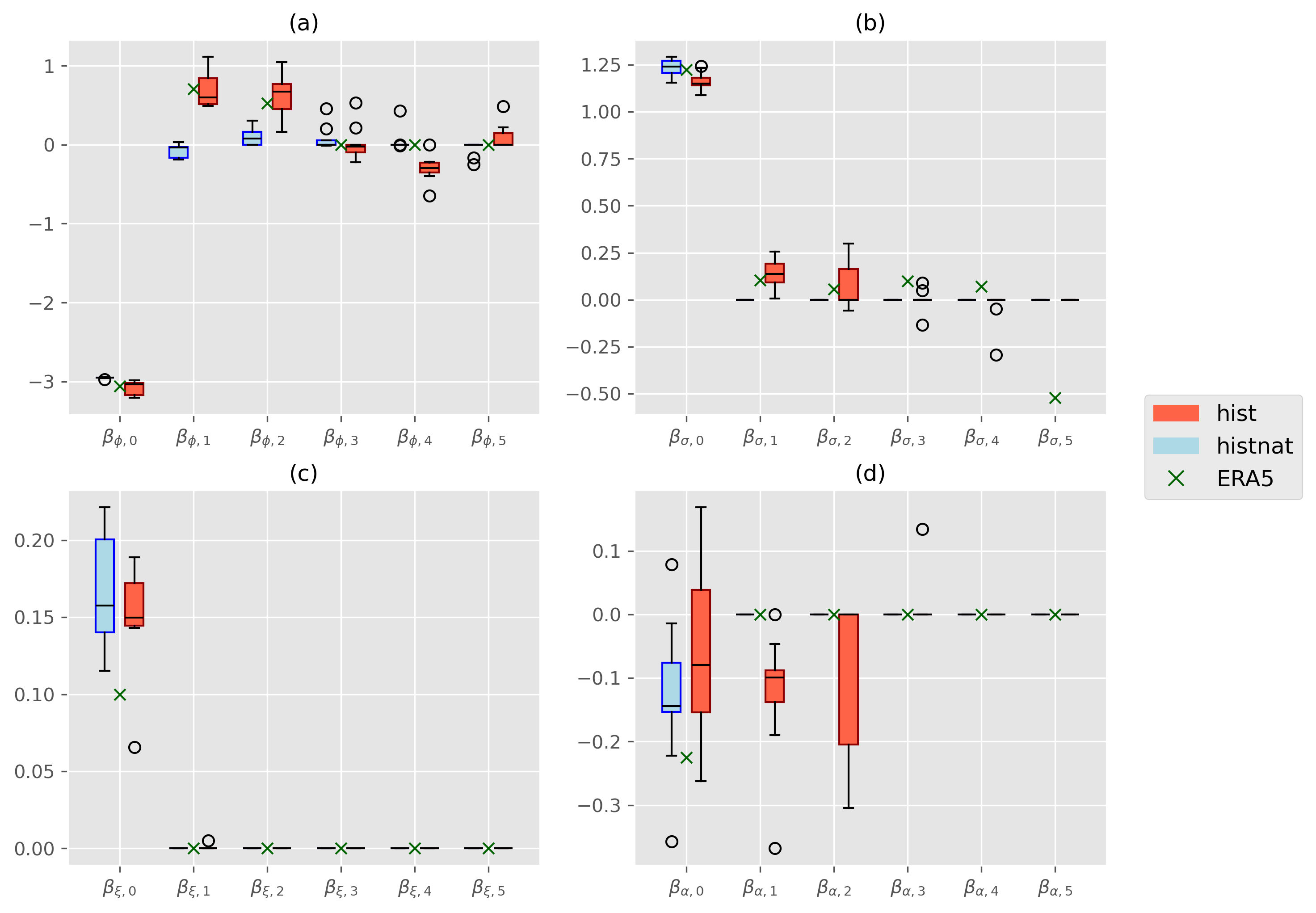}
    \caption{Same as Fig.~\ref{fig:parameter_estimates_model1} but for the northern European region. }
    \label{fig:parameter_estimates_model1_tasmax_reg_16}
\end{figure}

\begin{figure}[h]
    \centering
    \includegraphics[width=0.65\linewidth]{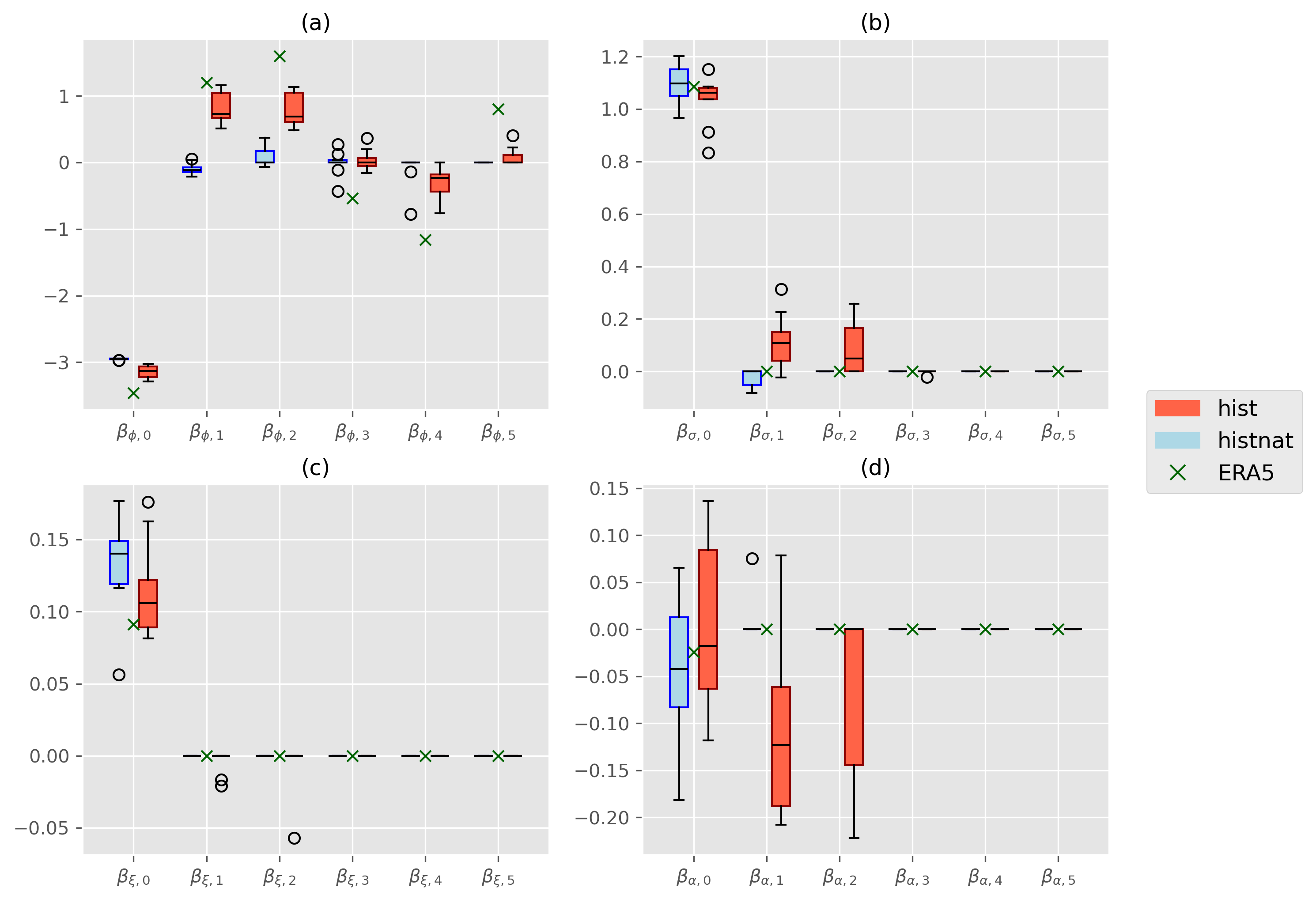}
    \caption{Same as Fig.~\ref{fig:parameter_estimates_model1} but for the central European region. }
    \label{fig:parameter_estimates_model1_tasmax_reg_17}
\end{figure}

The parameters for the northern and central European regions are estimated in a similar way to those for the southern European region, as shown in Sect.~\ref{sec:results}. For the model with a constant threshold, the estimated coefficients are quite similar across different regions (Figs.~\ref{fig:parameter_estimates_model1_tasmax_reg_16} and \ref{fig:parameter_estimates_model1_tasmax_reg_17}).

\begin{figure}[h]
    \centering
    \includegraphics[width=0.65\linewidth]{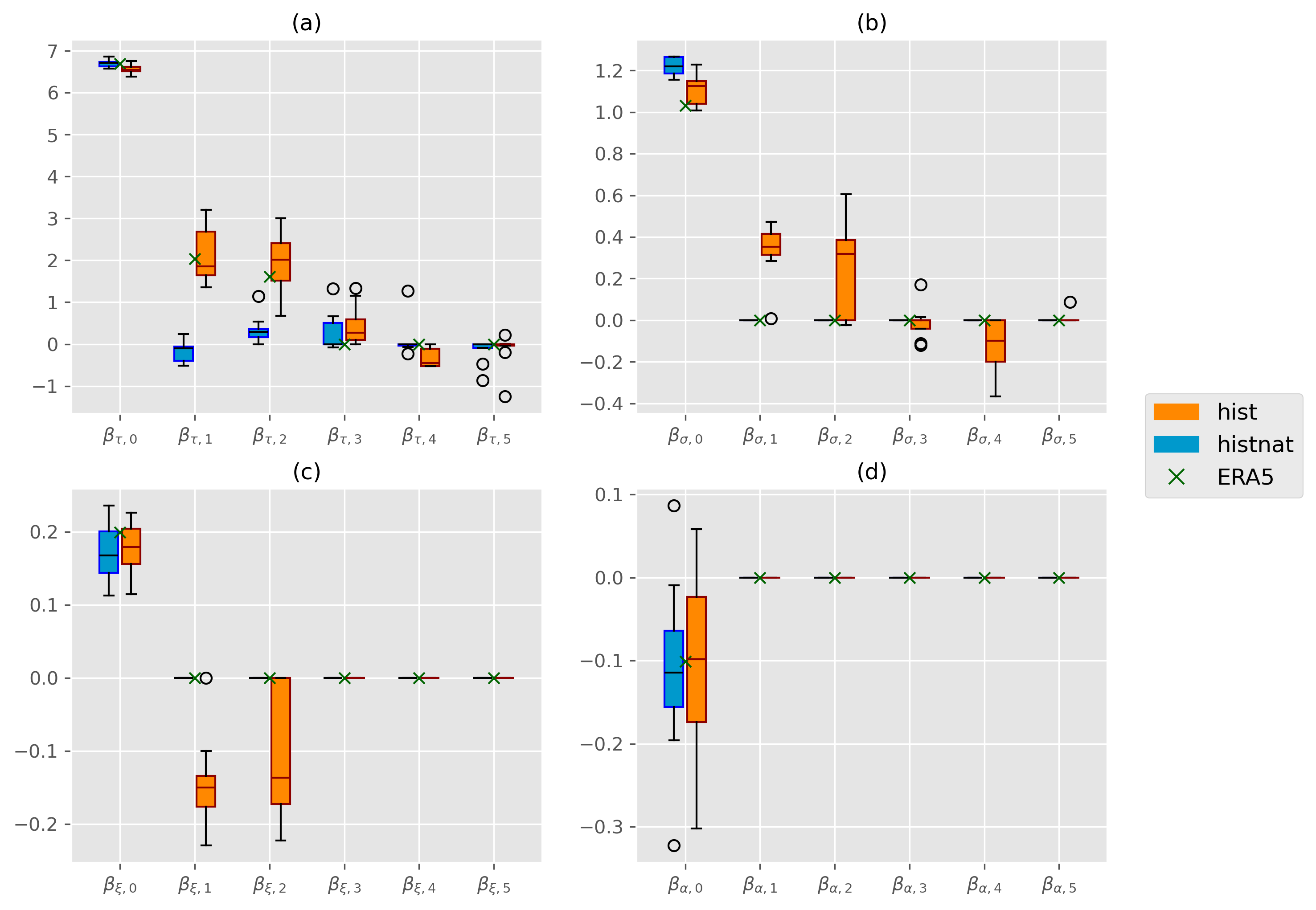}
    \caption{Same as Fig.~\ref{fig:parameter_estimates_model2} but for the northern European region. }
    \label{fig:parameter_estimates_model2_tasmax_reg_16}
\end{figure}

\begin{figure}[h]
    \centering
    \includegraphics[width=0.65\linewidth]{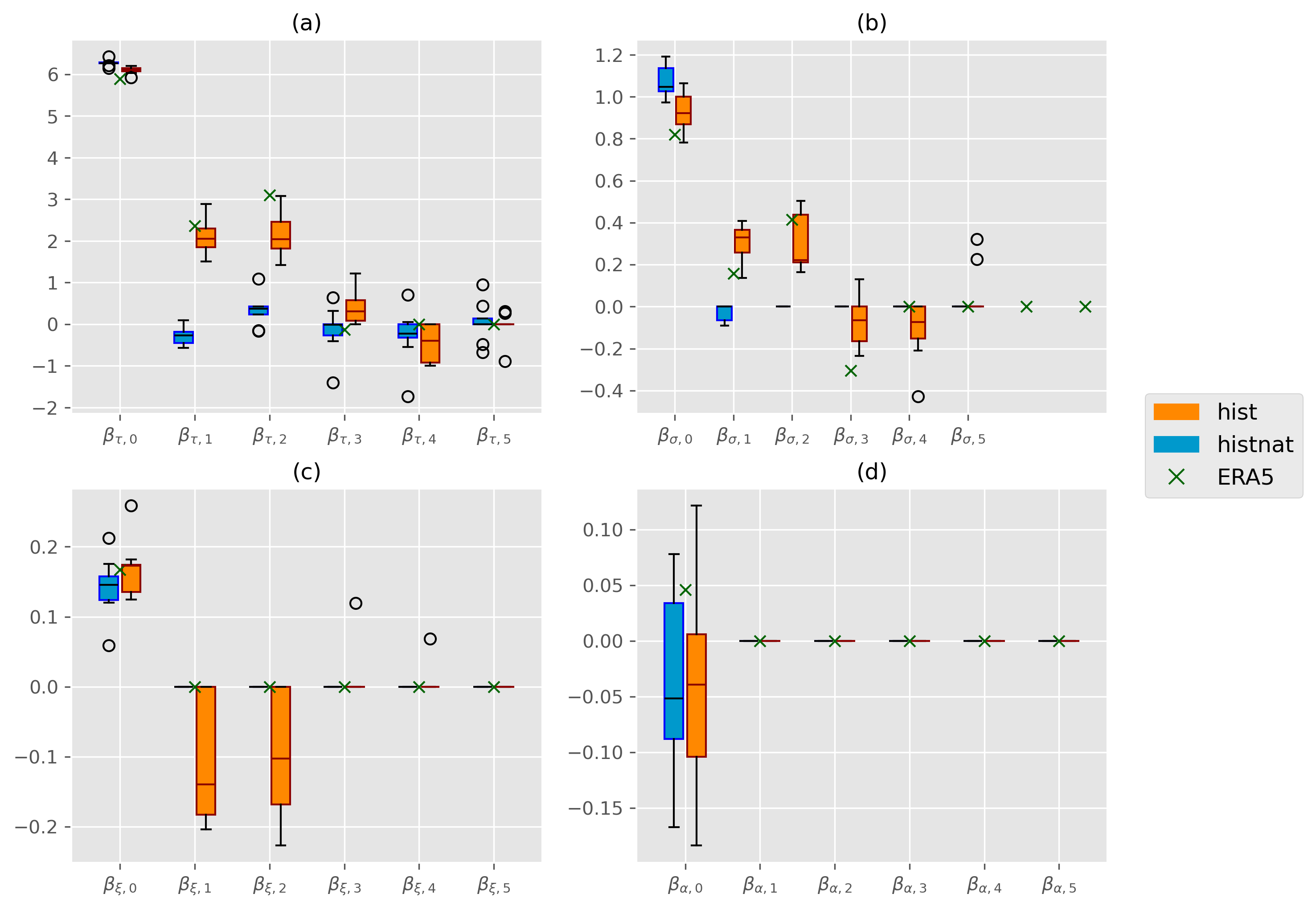}
    \caption{Same as Fig.~\ref{fig:parameter_estimates_model2} but for the central European region. }
    \label{fig:parameter_estimates_model2_tasmax_reg_17}
\end{figure}

As in the model with a constant threshold, the coefficients show high similarity for the time-varying threshold model (Figs.~\ref{fig:parameter_estimates_model2_tasmax_reg_16} and \ref{fig:parameter_estimates_model2_tasmax_reg_17}).

\subsection{Notation overview}
\label{sec:notation_overview}
Tables~\ref{tab:notation_data} and \ref{tab:notation_params} summarise the notation used
throughout this paper. The last column refers to the equation or section in which the
respective symbol is introduced.

\appendixtables

\begin{table*}[t]
\caption{Notation used in this study: indices, sample sizes, data, and distributions.}
\label{tab:notation_data}
\small
\begin{tabular}{p{0.14\textwidth}p{0.60\textwidth}p{0.20\textwidth}}
\tophline
Symbol & Meaning & Introduced in \\
\middlehline
\multicolumn{3}{l}{\textit{Indices}}\\[0.2em]
$i$ & Season (year) index; $i=1$ corresponds to the summer of 1940 & Eq.~(\ref{eqn:Ydef}) \\
$d$ & Day within a season & Eq.~(\ref{eqn:Ydef}) \\
$k$ & Extremal pattern and eigenvalue index (mode)& Sect.~\ref{sec:EPI} \\
$s$ & Scenario index, $s\in\{0,1\}$ & Sect.~\ref{sec:theory_attribution} \\
$r$ & Ensemble member index & Sect.~\ref{sec:model_estimation_with_constant_threshold} \\
$c$ & Climate model index & Sect.~\ref{sec:combining_lines_of_evidences} \\[0.5em]
\multicolumn{3}{l}{\textit{Sample sizes and model dimensions}}\\[0.2em]
$n_y$ & Number of seasons (summers 1940 to 2025) & Eq.~(\ref{eqn:Ydef}) \\
$n$ & Number of days per season ($n=92$, JJA) & Eq.~(\ref{eqn:Ydef}) \\
$q$ & Number of land grid points entering the TPDM & Sect.~\ref{sec:EPI} \\
$n_p$ & Number of leading extremal patterns retained ($n_p=10$) & Eq.~(\ref{eqn:EPI}) \\
$n_m$ & Number of climate models & Sect.~\ref{sec:combining_lines_of_evidences} \\
$n_r$ & Number of ensemble members of a climate model and scenario & Sect.~\ref{sec:model_estimation_with_constant_threshold} \\
$n_B$ & Number of bootstrap repetitions & Sect.~\ref{sec:uncertainty} \\
$K$ & Maximum degree of the Legendre polynomials ($K=5$) & Sect.~\ref{sec:nonstat} \\
$N$ & Most recent season included in the accumulated Bayes factor & Eq.~(\ref{Eq:LR}) \\[0.5em]
\multicolumn{3}{l}{\textit{Data}}\\[0.2em]
$\lambda_k$ & $k$-th eigenvalue of the TPDM & Sect.~\ref{sec:EPI} \\
$\eta_{k,i,d}$ & Principal component of pattern $k$ on day $d$ of season $i$ & Sect.~\ref{sec:EPI} \\
$y_{i,d}$ & EPI on day $d$ of season $i$; $y_{i,d}:=\mathrm{EPI}_{i,d}$ & Eq.~(\ref{eqn:EPI})\\
$\boldsymbol{y}^{(i)}$ & Time series of season $i$, $(y_{i,1},\ldots,y_{i,n})$ & Eq.~(\ref{eqn:Ydef}) \\
$\boldsymbol{Y}$ & Complete data set, $(\boldsymbol{y}^{(1)},\ldots,\boldsymbol{y}^{(n_y)})$ & Eq.~(\ref{eqn:Ydef}) \\
$\boldsymbol{Y}_{\mathrm{ERA5}}$, $\boldsymbol{y}^{(i)}_{\mathrm{ERA5}}$ & Corresponding quantities for the ERA5 reanalysis & Eq.~(\ref{eqn:probratio1}) \\[0.5em]
\multicolumn{3}{l}{\textit{Distributions and likelihoods}}\\[0.2em]
$F$ & Univariate marginal distribution function & Eq.~(\ref{eqn:marginal_gpd}) \\
$G$ & Joint distribution function of consecutive observations & Eq.~(\ref{eqn:full_tail}) \\
$l_1,l_2,l_3$ & Univariate, bivariate, and trivariate densities of the model & Eqs.~(\ref{eqn:markov_chain}), (\ref{eqn:markov_chain_2}) \\
$l(\cdot\,;\boldsymbol{\theta})$ & Likelihood of a season or of the complete data set & Eqs.~(\ref{eqn:markov_chain}), (\ref{eqn:full_likelihood}) \\
$V$ & Stable tail dependence function & Eq.~(\ref{eqn:full_tail}) \\
$V_1,V_2,V_{12}$ & Partial derivatives of $V$ with respect to its arguments & Eq.~(\ref{eqn:markov}) \\
$v_{i,d}$ & Transformed margin; $1/v_{i,d}$ is standard Fr\'echet distributed & Eq.~(\ref{eqn:margin}) \\
\bottomhline
\end{tabular}
\belowtable{}
\end{table*}

\begin{table*}[t]
\caption{Notation used in this study: model parameters, covariates, and attribution
quantities. Parameters are given in their general form with both indices; the assumption
under which the day index $d$ is dropped is stated in Eq.~(\ref{eqn:no_seasonal}).}
\label{tab:notation_params}
\small
\begin{tabular}{p{0.14\textwidth}p{0.60\textwidth}p{0.20\textwidth}}
\tophline
Symbol & Meaning & Introduced in \\
\middlehline
\multicolumn{3}{l}{\textit{Model parameters}}\\[0.2em]
$u_{i,d}$ & Threshold; constant $u$ in the first model, time-varying $u_i$ in the second & Eq.~(\ref{eqn:marginal_gpd}) \\
$\phi_{i,d}$ & Probability of exceeding the threshold; $\phi_i$ in the first model, $\phi=1-\tau$ in the second & Eq.~(\ref{eqn:marginal_gpd}) \\
$\sigma_{i,d}$, $\xi_{i,d}$ & Scale and shape parameters of the generalized Pareto distribution & Eq.~(\ref{eqn:marginal_gpd}) \\
$\alpha_{i,d}$ & Dependence parameter of the logistic model, attached to the transition from day $d$ to day $d+1$ & Eq.~(\ref{eqn:logistic_tail}) \\
$\tau$ & Quantile level of the quantile regression ($\tau=0.95$) & Sect.~\ref{sec:using_time_varying_threshold} \\
$u_{\mathrm{ERA5}}$, $u_{i,\mathrm{ERA5}}$ & Constant and time-varying threshold estimated from the ERA5 record & Eqs.~(\ref{eqn:probratio1}), (\ref{eqn:probratio2}) \\
$\boldsymbol{\theta}_1,\boldsymbol{\theta}_2$ & Marginal and dependence parameters; $\boldsymbol{\theta}=(\boldsymbol{\theta}_1,\boldsymbol{\theta}_2)$ & Eq.~(\ref{eqn:markov_chain}) \\[0.5em]
\multicolumn{3}{l}{\textit{Covariates and regression coefficients}}\\[0.2em]
$x_i$ & Season index rescaled to $[-1,1]$ & Eq.~(\ref{eqn:legendre}) \\
$\boldsymbol{g}_i$ & Legendre polynomials of degree $0$ to $K$, evaluated at $x_i$ & Eq.~(\ref{eqn:legendre}) \\
$\boldsymbol{h}_d$ & Sine and cosine of the day of the year (residual seasonal cycle) & Eq.~(\ref{eqn:seasonal_covariates}) \\
$\boldsymbol{\beta}_\phi$, $\boldsymbol{\beta}_\tau$ & Coefficients of the logistic and quantile regression & Eqs.~(\ref{eqn:logit}), (\ref{eqn:qr}) \\
$\boldsymbol{\beta}_\sigma$, $\boldsymbol{\beta}_\xi$, $\boldsymbol{\beta}_\alpha$ & Coefficients of the non-stationary scale, shape, and dependence parameters & Eqs.~(\ref{eqn:sigma_xi}), (\ref{eqn:alpha}) \\
$\boldsymbol{\delta}_\phi,\boldsymbol{\delta}_\sigma,\boldsymbol{\delta}_\xi,\boldsymbol{\delta}_\alpha$ & Coefficients of the residual seasonal cycle & Appendix~\ref{sec:seasonal_cycle} \\
$\rho_\tau$ & Check function of the quantile regression & Eq.~(\ref{eqn:check}) \\
$\mathcal{R}_\tau$ & Cost function minimised in the quantile regression & Eq.~(\ref{eqn:qs}) \\[0.5em]
\multicolumn{3}{l}{\textit{Attribution and combination of climate models}}\\[0.2em]
$m_0,m_1$ & Scenario without (HIST-NAT) and with (HIST) anthropogenic forcing & Sect.~\ref{sec:theory_attribution} \\
$p_0,p_1$ & Probability of an event under the two scenarios & Sect.~\ref{sec:theory_attribution} \\
$\mathrm{PR}$, $\mathrm{LR}$ & Probability ratio and likelihood ratio & Sect.~\ref{sec:theory_attribution} \\
$\mathrm{BF}$, $\mathrm{BF}(N)$ & Bayes factor, product of the seasonal likelihood ratios up to season $n_y$ (resp. $N$) & Eqs.~(\ref{eqn:BayesII}), (\ref{Eq:LR}) \\
$\hat{\mu}_c$, $\hat{\varsigma}_c$ & Estimate of the logarithmic likelihood ratio of climate model $c$ and its bootstrap standard deviation & Sect.~\ref{sec:combining_lines_of_evidences} \\
$w_c$ & Weight assigned to climate model $c$ & Sect.~\ref{sec:combining_lines_of_evidences} \\
$\varsigma_{\mathrm{nat}},\varsigma_{\mathrm{mod}},\varsigma_{\mathrm{tot}}$ & Natural, model, and total variability of the combined estimate & Sect.~\ref{sec:combining_lines_of_evidences} \\
$Q$ & Paule--Mandel statistic used to estimate $\varsigma_{\mathrm{mod}}$ & Sect.~\ref{sec:combining_lines_of_evidences} \\
\bottomhline
\end{tabular}
\belowtable{}
\end{table*}

\noappendix       

\codedataavailability{This study used a selection of ERA5 and CMIP6 data stored at DKRZ. ERA5 is freely available via the Copernicus data store, CMIP6 data can be retrieved from the ESGF data nodes. An implementation of the extremal pattern index is available via the ExtrPatt R-package (\url{https://CRAN.R-project.org/package=ExtrPatt}). The Python model code is provided, along with a Jupyter Notebook for the analysis and a Jupyter Notebook for visualisation. The extremal pattern index (EPI) output from the CMIP6 simulations and ERA5 reanalysis is available together with the code from Zenodo at \url{https://doi.org/10.5281/zenodo.20084245} \citep{pmeurer_2026_20084246}.}

\authorcontribution{All of the authors contributed to developing the idea for this study (conceptualisation). PF secured the funding of this work. PF and SB supervised the work. SvS provided support for the TPDM calculation. Data analysis and visualisation were performed by PM. PM also led the writing of the original draft. PF contributed to the original draft. All authors contributed to reviewing and editing the paper.}

\competinginterests{The authors declare that they have no conflict of interest.}

\begin{acknowledgements}
The authors thank Jonas Schröter and Erik Haufs for providing detailed background on the random effects model. We also thank the three anonymous reviewers for their valuable feedback, which has helped significantly improve the manuscript.
The flowchart and notation tables were generated using the Claude AI assistant (Anthropic).
The code used in this study was written entirely by the authors, without the use of AI tools. The version deposited at Zenodo differs from the original in that it has been revised to comply with the PEP 8 Python style guidelines. This reformatting was assisted by the Claude AI assistant (Anthropic).
\end{acknowledgements}

\financialsupport{This research was funded within the BMFTR project ClimXtreme II -- Module B under grant number FKZ 01LP2323A. This work used resources of the Deutsches Klimarechenzentrum (DKRZ) granted by its Scientific Steering Committee (WLA) under project ID bm1159.}

\bibliographystyle{copernicus}
\bibliography{Literatur.bib}

\end{document}